# 量子金鑰分配與 NIST 後量子密碼學標準演算法的混合方案應用於金鑰交換和數位簽章

Abel C. H. Chen
Information & Communications Security Laboratory,
Chunghwa Telecom Laboratories
Email: chchen.scholar@gmail.com; ORCID: 0000-0003-3628-3033

摘要

由於後量子密碼學(Post-Quantum Cryptography, PQC)演算法的安全性是建構在數學困難問題上，而量子金鑰分配(Quantum Key Distribution, QKD)的安全性是建構在量子物理原理上，各有優缺並且能夠互補。因此，開始有研究指出採用量子金鑰分配和後量子密碼學的混合方案，建構雙重保障。有鑑於此，本研究提出量子金鑰分配與 NIST 後量子密碼學標準演算法的混合方案，包含量子金鑰分配與後量子密碼學混合方案的金鑰交換、量子金鑰分配與後量子密碼學混合方案的數位簽章。在量子金鑰分配與後量子密碼學混合方案的金鑰交換設計上，本研究主要結合模晶格金鑰封裝機制(Module-Lattice-Based Key-Encapsulation Mechanism, ML-KEM)和 BB84協定、E91協定來建立金鑰交換協定。在量子金鑰分配與後量子密碼學混合方案的數位簽章設計上，本研究主要在模晶格數位簽章演算法(Module-Lattice-Based Digital Signature Algorithm, ML-DSA)、無狀態雜湊數位簽章演算法(Stateless Hash-Based Digital Signature Algorithm, SLH-DSA)基礎上建構簽章重構值，並且搭配 BB84協定、E91協定傳輸確認碼後得以驗證簽章值。在實驗中，本研究驗證量子金鑰分配與後量子密碼學混合方案的金鑰交換所產製的共享秘密金鑰(shared secret key)，從熵(entropy)、獨立且同分布(Independent and Identically Distributed, IID)等面向來驗證性能。此外，本研究也從計算時間驗證量子金鑰分配與後量子密碼學混合方案的金鑰交換、量子金鑰分配與後量子密碼學混合方案的數位簽章。

# 量子金鑰分配與 NIST 後量子密碼學標準演算法的混合方案應用於金鑰交換和數位簽章

## 摘要

由於後量子密碼學(Post-Quantum Cryptography, PQC)演算法的安全性是建構在數學困難問題上，而量子金鑰分配(Quantum Key Distribution, QKD)的安全性是建構在量子物理原理上，各有優缺並且能夠互補。因此，開始有研究指出採用量子金鑰分配和後量子密碼學的混合方案，建構雙重保障。有鑑於此，本研究提出量子金鑰分配與 NIST 後量子密碼學標準演算法的混合方案，包含量子金鑰分配與後量子密碼學混合方案的金鑰交換、量子金鑰分配與後量子密碼學混合方案的數位簽章。在量子金鑰分配與後量子密碼學混合方案的金鑰交換設計上，本研究主要結合模晶格金鑰封裝機制(Module-Lattice-Based Key-Encapsulation Mechanism, ML-KEM)和 BB84協定、E91協定來建立金鑰交換協定。在量子金鑰分配與後量子密碼學混合方案的數位簽章設計上，本研究主要在模晶格數位簽章演算法(Module-Lattice-Based Digital Signature Algorithm, ML-DSA)、無狀態雜湊數位簽章演算法(Stateless Hash-Based Digital Signature Algorithm, SLH-DSA)基礎上建構簽章重構值，並且搭配 BB84協定、E91協定傳輸確認碼後得以驗證簽章值。在實驗中，本研究驗證量子金鑰分配與後量子密碼學混合方案的金鑰交換所產製的共享秘密金鑰(shared secret key)，從熵(entropy)、獨立且同分布(Independent and Identically Distributed, IID)等面向來驗證性能。此外，本研究也從計算時間驗證量子金鑰分配與後量子密碼學混合方案的金鑰交換、量子金鑰分配與後量子密碼學混合方案的數位簽章。

## 1. 前言

美國國家標準暨技術研究院(National Institute of Standards and Technology, NIST)已經在 2024 年 8 月制定出後量子密碼學(Post-Quantum Cryptography, PQC)標準演算法，明確規範金鑰封裝機制(Key Encapsulation Mechanism, KEM)[1]和數位簽章演算法(Digital Signature Algorithm, DSA)[2]-[3]的相關流程和參數，可以在數學困難問題上保障安全性[4]，為全球資訊安全領域做出巨大的貢獻。除此之外，量子計算技術逐漸成熟，量子金鑰分配(Quantum Key Distribution, QKD)也開始被採用作為建構在量子物理原理上的金鑰封裝機制，可以傳送共享秘密金鑰(shared secret key)[5]。因此，GSM 協會(Groupe Speciale Mobile Association, GSMA)在 2024 年年底時發佈白皮書「IG.18 Opportunities and Challenges for Hybrid (QKD and PQC) Scenarios」，內容提及未來在 6G 的核心網路設備之間的連線可以建構在量子金鑰分配與後量子密碼學演算法的混合方案，以提供雙重保障[6]。美國國家標準暨技術研究院負責後量子密碼學專案主要成員 Yi-Kai Liu 和 Dustin Moody 也於 2024 年在《Physical Review Applied》發表重要的研究成果「Post-quantum cryptography and the quantum future of cybersecurity」，內容指出量子金鑰分配與後量子密碼學演算法建構在不同的安全特性，具有互補的效果[7]。綜合上述，量子金鑰分配與後量子密碼學演算法的混合方案是重要的研究方向之一。

隨著後量子密碼學標準化的發展，國際上也陸續開始制定 RSA 密碼學、橢圓曲線密碼學(Elliptic Curve Cryptography, ECC)與後量子密碼學的混合方案，讓安全性可以建構在不同的數學困難問題上，達到多重安全保障。此外，部分混合方案在設計上也考慮從 RSA 密碼學、橢圓曲線密碼學遷移到後量子密碼學過渡期的方案，有很多傑出且非常值得借鑑的設計。在金鑰交換設計上，密碼學資深專家們 Douglas Stebila 教授、Scott Fluhrer 工程師、Shay Gueron 教授提出重要的研究成果「Hybrid key exchange in TLS 1.3」，為多個密碼學方法混合方案的金鑰交換協定提供明確且高效能的設計[8]。在數位簽章設計上，資訊安全資深專家們 Mike Ounsworth 工程師、John Gray 工程師作為主要貢獻者之一，提出多種不同的混合憑證方案，包含 Composite [9]、Catalyst [10]、Chameleon [11]等，讓同一張憑證內可以包含多個密碼學方法的簽章。

為提供多重安全保障，本研究在現行的混合方案基礎上提出量子金鑰分配與 NIST 後量子密碼學標準演算法的混合方案，並且應用在金鑰交換和數位簽章。其中，在量子金鑰分配與後量子密碼學混合方案的金鑰交換設計上，本研究主要結合模晶格金鑰封裝機制(Module-Lattice-Based Key-Encapsulation Mechanism, ML-KEM)[1]和 BB84 協定[12]、E91 協定[13]來建立金鑰交換協定。在量子金鑰分配與後量子密碼學混合方案的數位簽章設計上，本研究主要在模晶格數位簽章演算法(Module-Lattice-Based Digital Signature Algorithm, ML-DSA)[2]、無狀態雜湊數位簽章演算法(Stateless Hash-Based Digital Signature Algorithm, SLH-DSA)[3]基礎上建構簽章重構值，並且搭配 BB84 協定[12]、E91 協定[13]傳輸確認碼後得以驗證簽章值。本研究主要貢獻條列如下：

- 本研究提出量子金鑰分配與後量子密碼學混合方案的金鑰交換，結合量子金鑰分配得到的隨機數和模晶格金鑰封裝機制得到的隨機數，再代入金鑰衍生函數(Key Derivation Function, KDF)[14]得到共享秘密金鑰。
- 本研究提出量子金鑰分配與後量子密碼學混合方案的數位簽章，運用模晶格數位簽章演算法和無狀態雜湊數位簽章演算法產製簽章及其簽章重構值，結合量子金鑰分配傳輸確認碼後得以驗證簽章值。

- 為驗證金鑰交換得到的共享秘密金鑰之隨機性，本研究參考 NIST SP 800-90B 所述驗證方法[15]分析是否符合隨機位元熵(entropy)和隨機位元獨立且同分布(Independent and Identically Distributed, IID)。
- 本研究亦驗證每一個方法的計算效率，以作為後續部署金鑰交換和數位簽章應用時參考。

本文分為六個章節。第 2 節介紹現行的金鑰交換協定和多密碼學混合的數位簽章。第 3 節提出本研究設計的量子金鑰分配與後量子密碼學混合方案，包含金鑰交換和數位簽章兩個應用面向。第 4 節說明本研究使用的量子金鑰分配具體作法，並且分別深入介紹 BB84 協定和 E91 協定。第 5 節對本研究提出的方法進行驗證，分別從共享秘密金鑰熵驗證、獨立且同分布、計算效率等面向驗證。最後，第 6 節總結本研究主要發現，並且討論未來研究方向。

## 2. 混合方案相關研究

本節將依序介紹現行的單一密碼學的金鑰交換協定、基於橢圓曲線密碼學與後量子密碼學混合方案的金鑰交換協定、以及基於橢圓曲線密碼學與後量子密碼學混合方案的數位簽章。

### 2.1 單一密碼學的金鑰交換協定

本節將說明單一密碼學的金鑰交換協定，並假設 Alice 為發起端，Bob 為接收端，雙方在協商得可以得共享秘密金鑰。以下依序說明基於橢圓曲線密碼學的金鑰交換協定、基於模晶格金鑰封裝機制的金鑰交換協定、以及基於量子金鑰分配協定的金鑰交換協定。

#### 1) 基於橢圓曲線密碼學的金鑰交換協定

現行的傳輸層安全性協定(Transport Layer Security, TLS)主要採用基於橢圓曲線密碼學的金鑰交換協定[16]，其運作方式如圖 1 所示。為了提升安全性，每次連線時都將產製暫時(ephemeral)橢圓曲線密碼學金鑰對。首先，由 Alice 產製 Alice 的暫時橢圓曲線密碼學金鑰對(私鑰為$p_A$，公鑰為$P_A$，如公式(1)所示)，並且在 ClientHello 訊息中代入 Alice 的暫時橢圓曲線密碼學公鑰$P_A$傳送給 Bob。當 Bob 收到 ClientHello 訊息後將產製 Bob 的暫時橢圓曲線密碼學金鑰對(私鑰為$p_B$，公鑰為$P_B$，如公式(2)所示)，並且在 ServerHello 訊息中代入 Bob 的暫時橢圓曲線密碼學公鑰$P_B$回傳給 Alice。後續雙方可以執行橢圓曲線 Diffie-Hellman (Elliptic-Curve Diffie–Hellman, ECDH)，用自身的私鑰乘上對方的公鑰取得橢圓曲線點$P_C$ (如公式(3)所示)，並以橢圓曲線點$P_C$的 $x$ 座標值作為共享秘密金鑰 $r_1$。其中，$G$ 為橢圓曲線基點。

$$P_A = p_A G. \tag{1}$$

$$P_B = p_B G. \tag{2}$$

$$P_C = p_A P_B = p_B P_A = p_A p_B G. \tag{3}$$

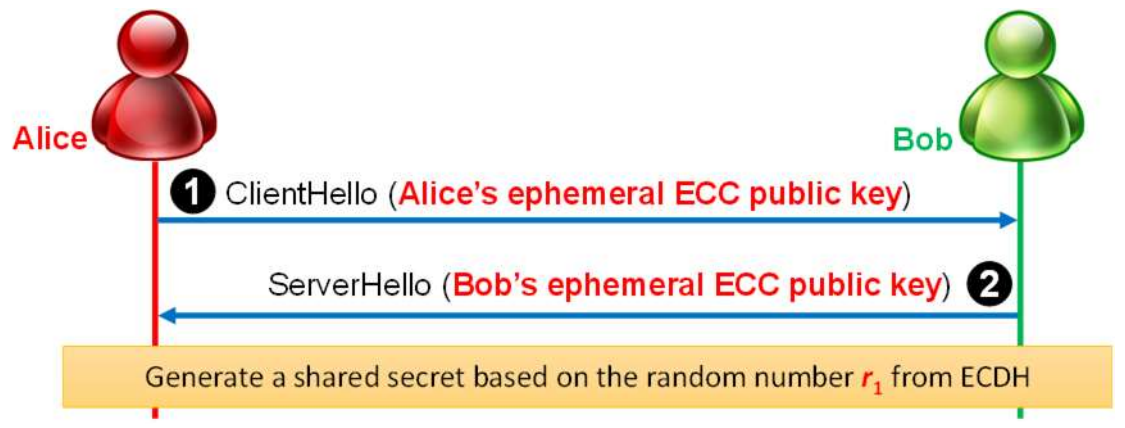

圖 1 基於橢圓曲線密碼學金鑰交換協定

#### 1) 基於模晶格金鑰封裝機制的金鑰交換協定

傳輸層安全性協定採用基於模晶格金鑰封裝機制的金鑰交換協定[8]作法如圖 2 所示。首先，由 Alice 產製 Alice 的暫時模晶格金鑰封裝機制金鑰對，並且在 ClientHello 訊息中代入 Alice 的暫時模晶格金鑰封裝機制公鑰傳送給 Bob。當 Bob 收到 ClientHello 訊息後將運用 Alice 的暫時模晶格金鑰封裝機制公鑰加密共享秘密金鑰 $r_2$ 得到密文，再把密文(即金鑰封裝值)放到 ServerHello 訊息中和回傳給 Alice。後續 Alice 收到 ServerHello 訊息後，可以用 Alice 的暫時模晶格金鑰封裝機制私鑰解密密文得到共享秘密金鑰 $r_2$。

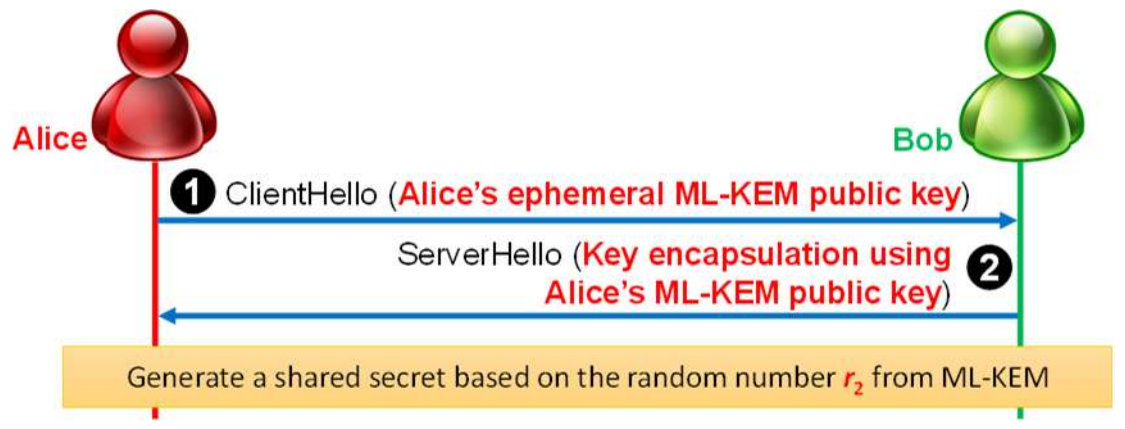

圖 2 基於模晶格金鑰封裝機制的金鑰交換協定

#### 1) 基於量子金鑰分配協定的金鑰交換協定

在基於量子金鑰分配協定的金鑰交換協定，首先將由 Alice 和 Bob 雙方建立量子通道(quantum channel)傳送具有量子態，並且根據各自己的控制訊號進行操作後量測，再通訊經典通道(classical channel)傳送控制訊號和比對彼此的控制訊號取得合適的隨機位元值作為共享秘密金鑰 $r_3$，如圖 3 所示。

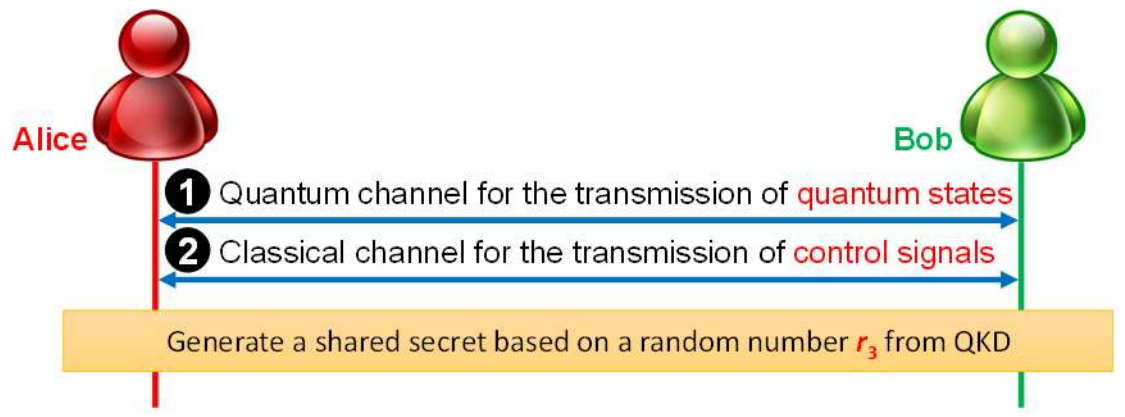

圖 3 基於量子金鑰分配協定的金鑰交換協定

其中，知名的量子金鑰分配協定包含有 BB84 協定[12]、E91 協定[13]等。本研究將於第 4 節詳細介紹本研究所採用的 BB84 協定和 E91 協定，並且分別於附錄 A 和附錄 B 詳細說明 BB84 協定和 E91

協定的安全性。討論在雙方採用各種控制訊號組合下的量子態變化，並且在雙方採用的控制訊號不同時可以有均勻疊加態，從而保障量子金鑰分配協定的安全性。

### 2.2 基於橢圓曲線密碼學與後量子密碼學混合方案的金鑰交換協定

網際網路工程任務組(Internet Engineering Task Force, IETF)草案(draft)「Hybrid key exchange in TLS 1.3」中已經規範 RSA 密碼學、橢圓曲線密碼學與後量子密碼學混合方案的金鑰交換協定，圖 4 為基於橢圓曲線密碼學與後量子密碼學混合方案的金鑰交換協定。其核心精神主要在於合併第 2.1.1 節和第 2.1.2 節所描述的 ClientHello 訊息內容和 ServerHello 訊息內容。

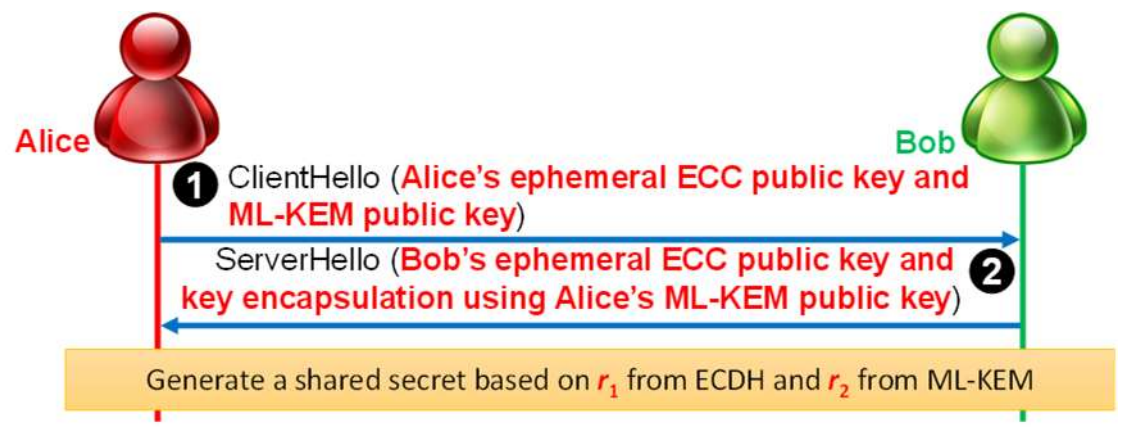

**圖 4 基於橢圓曲線密碼學與後量子密碼學混合方案的金鑰交換協定**

首先，Alice 產製 Alice 的暫時橢圓曲線密碼學金鑰對和 Alice 的暫時模晶格金鑰封裝機制金鑰對，並且在 ClientHello 訊息中代入 Alice 的暫時橢圓曲線密碼學公鑰和暫時模晶格金鑰封裝機制公鑰傳送給 Bob。之後，當 Bob 收到 ClientHello 訊息後將產製 Bob 的暫時橢圓曲線密碼學金鑰對，並且運用 Alice 的暫時模晶格金鑰封裝機制公鑰加密共享秘密金鑰 $r_2$ 得到金鑰封裝值(即密文)，再把 Bob 的暫時橢圓曲線密碼學公鑰和金鑰封裝值放到 ServerHello 訊息中和回傳給 Alice。後續 Alice 收到 ServerHello 訊息後，可以用 Alice 的暫時模晶格金鑰封裝機制私鑰解密密文得到隨機數 $r_2$。以及 Alice 和 Bob 雙方可以通過 ECDH 和公式(3)計算得到隨機數 $r_1$。

在取得隨機數 $r_1$ 和隨機數 $r_2$ 後，再運用已被正式認可的(approved)金鑰衍生函數來進行計算得到共享秘密金鑰。其中，已被正式認可的金鑰衍生函數可參考 NIST SP 800-108 Rev. 1，例如：基於 Keccak 訊息鑑別碼(Keccak-based Message Authentication Code, KMAC)的金鑰衍生函數[14]。

### 2.3 基於橢圓曲線密碼學與後量子密碼學混合方案的數位簽章

目前在 RSA 密碼學、橢圓曲線密碼學與後量子密碼學混合方案的數位簽章發展上，主要先應用在 X.509 混合憑證。其中，Mike Ounsworth 工程師、John Gray 工程師等多位資訊安全資深專家提出 Composite [9]、Catalyst [10]、Chameleon [11]三種不同的混合憑證方案。

圖 5 為基於 Catalyst 方案的橢圓曲線密碼學與模晶格數位簽章演算法 X.509 混合憑證。為簡化說明，本節以 Alice 的憑證為例，在憑證 Subject Public Key Info 欄位中的 alg 欄位放置 ML-DSA-44 OID 和 key 欄位放置憑證擁有者(即 Alice)的 ML-DSA-44 公鑰，以及在 Sig 欄位放置憑證簽發者(即憑證中心(Certificate Authority, CA))對這張憑證的 ML-DSA-44 簽章值。另外，Catalyst 方案的 X.509 混合憑證的作法是在 Extensions 欄位中增加 altAlg 欄位放置橢圓曲線數位簽章演算法(Elliptic Curve Digital Signature Algorithm, ECDSA) NIST P-256 OID、altKey 欄位放置憑證擁有者的 ECDSA NIST P-256 公鑰、以及 altSig 欄位放置憑證簽發者對這張憑證的 ECDSA NIST P-256 簽章值。

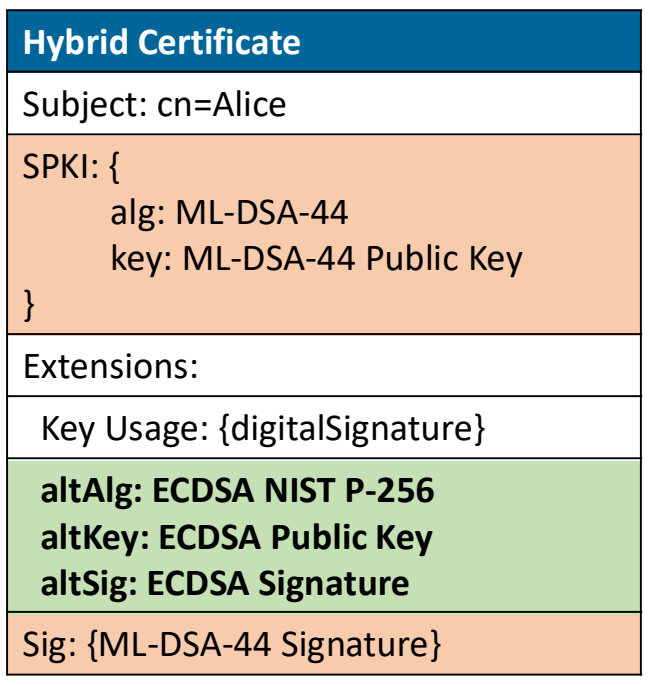

**圖 5 基於 Catalyst 方案的橢圓曲線密碼學與模晶格數位簽章演算法 X.509 混合憑證**

當其他設備取得這張 X.509 混合憑證時，將運用憑證中心的 ML-DSA-44 公鑰驗證 Sig 欄位裡的 ML-DSA-44 簽章值，以及運用憑證中心的 ECDSA NIST P-256 公鑰驗證 altSig 欄位裡的 ECDSA NIST P-256 簽章值。當兩個簽章值都驗證通過，並且憑證其他資訊(如：效期和金鑰用途(Key Usage))也驗證通過後，則可以信任此憑證。

## 3. 本研究提出的量子金鑰分配與後量子密碼學混合方案

本節將詳細介紹本研究提出的量子金鑰分配與後量子密碼學混合方案，第 3.1 節說明本研究設計的量子金鑰分配與後量子密碼學混合方案的金鑰交換，第 3.2 節說明本研究設計的量子金鑰分配與後量子密碼學混合方案的數位簽章。

### 3.1 量子金鑰分配與後量子密碼學混合方案的金鑰交換

在金鑰交換的應用上，本研究設計了兩個方法：方法(1)結合量子金鑰分配與後量子密碼學，方法(2)結合量子金鑰分配、後量子密碼學、橢圓曲線密碼學混合方案，分述如下。

**2)** *方法(1)：基於量子金鑰分配與後量子密碼學混合方案的金鑰交換協定*

本研究提出的基於量子金鑰分配與後量子密碼學混合方案的金鑰交換協定主要建構在第 2.1.2 節和第 2.1.3 節的基礎上，如圖 6 所示。

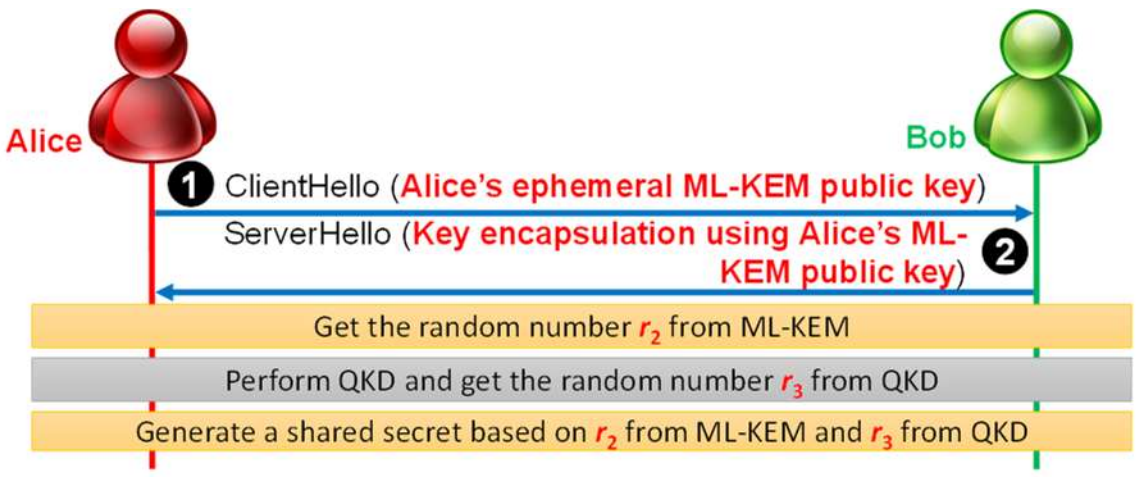

**圖 6 基於量子金鑰分配與後量子密碼學混合方案的金鑰交換協定**

首先，先執行基於模晶格金鑰封裝機制的金鑰交換協定，由 Alice 和 Bob 通過第 2.1.2 節描述的流程得到隨機數 $r_2$。其中，Alice 在 ClientHello 訊息中代入 Alice 的暫時模晶格金鑰封裝機制公鑰傳送給 Bob；由 Bob 加密隨機數 $r_2$ 得到金鑰封裝值，並且在 ServerHello 訊息中代入金鑰封裝值後回傳給 Alice。然後，Alice 用 Alice 的暫時模晶格金鑰封裝機制私鑰解密密文得到共享秘密金鑰 $r_2$。

在完成基於模晶格金鑰封裝機制的金鑰交換協定後，再執行基於量子金鑰分配協定的金鑰交換協定，由 Alice 和 Bob 通過第 2.1.3 節描述的流程得到隨機數 $r_3$，詳細過程請見第 4 節。

在取得隨機數 $r_2$ 和隨機數 $r_3$ 後，再運用已被正式認可的金鑰衍生函數來進行計算得到共享秘密金鑰。其中，本研究設定隨機數 $r_3$ 作為金鑰，而隨機數 $r_2$ 作為訊息，以及指定輸出長度為 32 bytes，代入基於 Keccak 訊息鑑別碼(KMAC)的金鑰衍生函數[14]產生 32-byte 長度的隨機數作為共享秘密金鑰。

**3)** *方法(2)：基於量子金鑰分配、後量子密碼學、橢圓曲線密碼學混合方案的金鑰交換協定*

本研究提出的基於量子金鑰分配、後量子密碼學、橢圓曲線密碼學混合方案的金鑰交換協定主要建構在第 2.2 節和第 2.1.3 節的基礎上，如圖 7 所示。

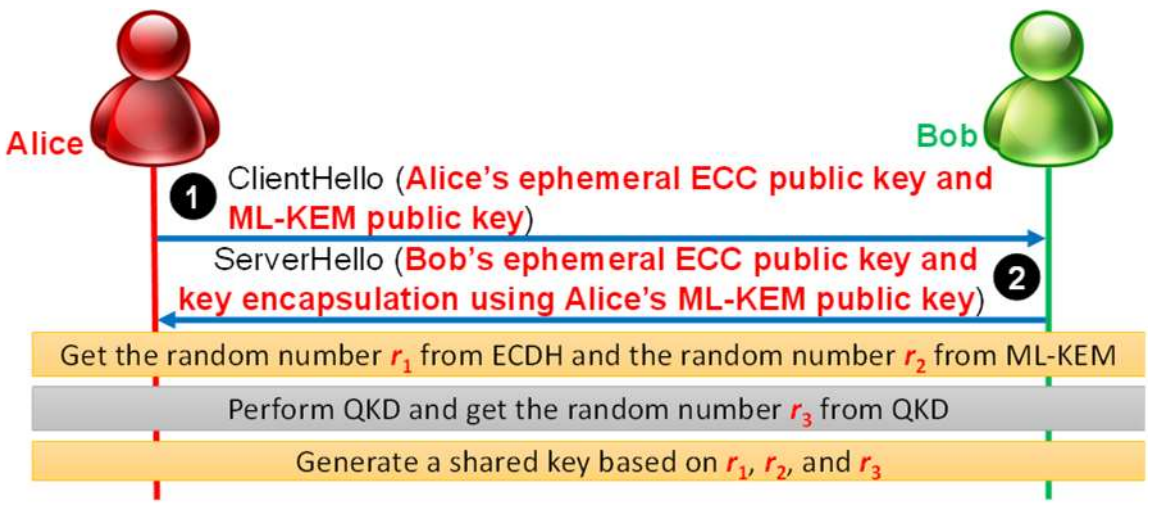

**圖 7 基於量子金鑰分配、後量子密碼學、橢圓曲線密碼學混合方案的金鑰交換協定**

首先，先執行基於橢圓曲線密碼學與後量子密碼學混合方案的金鑰交換協定，由 Alice 和 Bob 通過第 2.2 節描述的流程得到隨機數 $r_1$ 和隨機數 $r_2$。其中，Alice 在 ClientHello 訊息中代入 Alice 的暫時橢圓曲線密碼學公鑰和 Alice 的暫時模晶格金鑰封裝機制公鑰傳送給 Bob；由 Bob 加密隨機數 $r_2$ 得到金鑰封裝值，並且在 ServerHello 訊息中代入 Bob 的暫時橢圓曲線密碼學公鑰和金鑰封裝值後回傳給 Alice。然後，Alice 用 Alice 的暫時模晶格金鑰封裝機制私鑰解密密文得到共享秘密金鑰 $r_2$。以及 Alice 和 Bob 雙方可以通過 ECDH 和公式(3)計算得到隨機數 $r_1$。

在完成基於模晶格金鑰封裝機制的金鑰交換協定後，再執行基於量子金鑰分配協定的金鑰交換協定，由 Alice 和 Bob 通過第 2.1.3 節描述的流程得到隨機數 $r_3$，詳細過程請見第 4 節。

在取得隨機數 $r_1$、隨機數 $r_2$、隨機數 $r_3$ 後，再運用已被正式認可的金鑰衍生函數來進行計算得到共享秘密金鑰。其中，本研究設定隨機數 $r_3$ 作為金鑰，而訊息內容是所 $r_1||r_2$ 組成，以及指定輸出長度為 32 bytes，代入基於 Keccak 訊息鑑別碼(KMAC)的金鑰衍生函數[14]產生 32-byte 長度的隨機數作為共享秘密金鑰。

## 3.2 量子金鑰分配與後量子密碼學混合方案的數位簽章

本節以 X.509 混合憑證的簽章和驗章為例來說明本研究提出的基於量子金鑰分配與後量子密碼學混合方案的數位簽章。

在 X.509 混合憑證的格式設計上，主要參考第 2.3 節的方法，在 Extensions 欄位中增加 QKDInfo 欄位放置憑證簽發者(即憑證中心)的量子金鑰分配所需的量子通道和經典通道資訊，如圖 8 所示。並且為了更進一步縮小憑證中的簽章值長度，本研究在 Sig 欄位僅放置一個 32-byte 長度的簽章請求碼(Request Code for Signature, RCS) $r_4$，簽章和驗章的詳細流程(如圖 9 所示)分述如下。

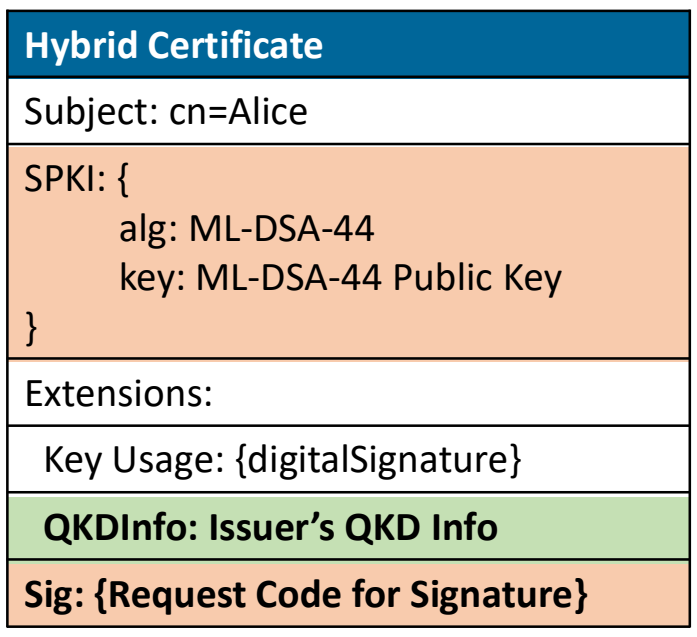

**圖 8 量子金鑰分配與後量子密碼學混合方案的 X.509 混合憑證**

**1)** **簽章流程**

假設憑證中心的數位簽章演算法採用 ML-DSA-44，並且已經收到可信任的註冊中心(Registration Authority, RA)轉送過來的 Alice 憑證簽章請

求(Certificate Signing Request, CSR)，並且包含 Alice 資訊及其 ML-DSA-44 公鑰。

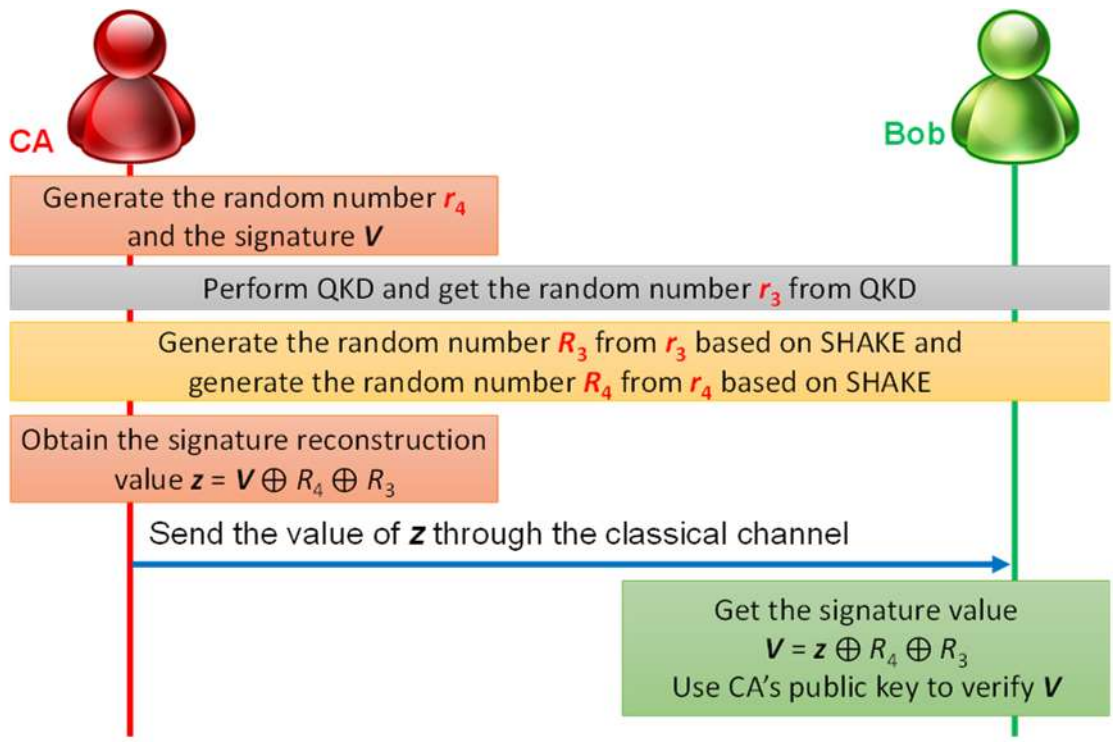

**圖 9 量子金鑰分配與後量子密碼學混合方案的數位簽章**

憑證中心將產製 Alice 憑證，在 SPKI 欄位放置 Alice 的 ML-DSA-44 公鑰，而 QKDInfo 欄位放置憑證中心的量子金鑰分配所需的量子通道和經典通道資訊，並且對該憑證的待簽章資料(To-Be-Signed-Data)產製 ML-DSA-44 簽章值 $V$，以及產製一個 32-byte 長度的隨機數 $r_4$ 作為簽章請求碼放置在憑證 Sig 欄位。

#### 2) 驗章流程

當 Bob 取得 Alice 憑證，欲驗證 Alice 憑證的有效性時，可以通過憑證中的 QKDInfo 欄位資訊與憑證中心進行量子金鑰分配並雙方各自取得 32-byte 長度的隨機數 $r_3$ (如第 2.1.3 節所述)。

Bob 和憑證中心可以各自分別把隨機數 $r_3$ 和隨機數 $r_4$ 帶入 SHAKE 演算法，並且指定輸出長度為數位簽章演算法對應的簽章值長度(例如，ML-DSA-44 的簽章值長度為 2420 bytes)。通過 SHAKE 演算法，可以把隨機數 $r_3$ 擴展為 2420-byte 長度的隨機數 $R_3$，以及把隨機數 $r_4$ 擴展為 2420-byte 長度的隨機數 $R_4$。

然後，憑證中心運用公式(4)計算簽章重構值(Signature Reconstruction Value, SRV) $z$，並且用經典通道傳送簽章重構值 $z$ 給 Bob。Bob 收到簽章重構值 $z$ 後運用公式(5)計算得到 ML-DSA-44 簽章值 $V$。之後，Bob 取得 Alice 憑證的待簽章資料(To-Be-Signed-Data)和憑證中心的 ML-DSA-44 公鑰，對 ML-DSA-44 簽章值 $V$ 進行驗章。

$$z = V \oplus R_4 \oplus R_3. \quad (4)$$

$$V = z \oplus R_4 \oplus R_3. \quad (5)$$

## 4. 本研究使用的量子金鑰分配

本節將深入說明本研究採用的 BB84 協定和 E91 協定，並且將分別在第 4.1 節和第 4.2 節說明本研究在這兩個協定設定的控制訊號，而量子態證明則在附錄 A 和附錄 B 討論。第 4.3 節將說明本研究對協商後訊息的處理方法，以產製固定 $n$ 位元長度的訊息。

### 4.1 BB84協定

在 BB84 協定[12]基礎上，當 Alice 和 Bob 想協商$n$個位元的訊息(例如：$n$個位元的共享秘密金鑰)，則需要用到$2 \times n$個量子位元來傳輸。假設 Alice 為發起端，而 Bob 為接收端，BB84 協定詳細流程如圖 10 所示，具體步驟描述如下。

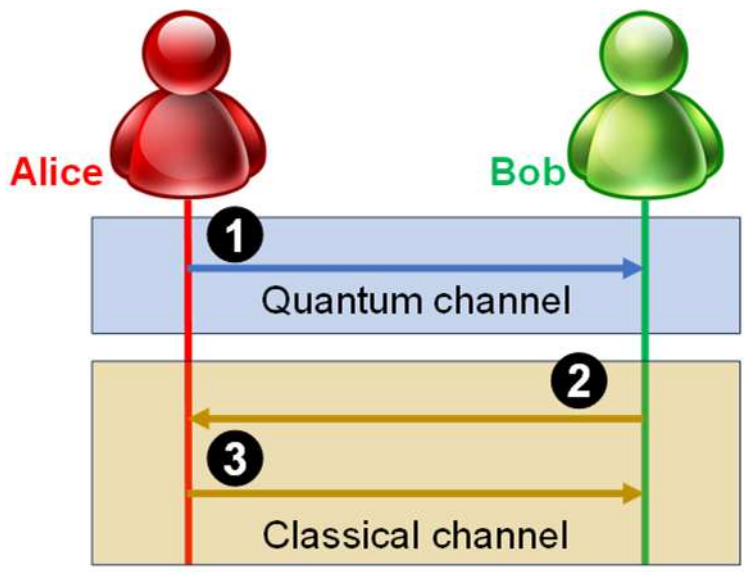

**圖 10 BB84 協定**

Alice 首先依據下列步驟產生和傳送量子訊號：

(a). Alice 隨機產生$2 \times n$個量子位元值，表示式為$Q_A = \{q_{A,2n-1}, q_{A,2n-2}, \dots, q_{A,1}, q_{A,0}\}$；其中，每個量子位元初始值皆為$|0\rangle$，每個量子位元有$\frac{1}{2}$的機率會被操作 Pauli-X 閘變換為$|1\rangle$。

(b). Alice 隨機產生$2 \times n$個控制訊號，表示式為$A = \{a_{2n-1}, a_{2n-2}, \dots, a_1, a_0\}$；其中，第 $i$ 個控制訊號$a_i$決定第 $i$ 個量子位元$q_{A,i}$是否操作 Hadamard gate (H-gate)，每個量子位元有$\frac{1}{2}$的機率會被操作 Hadamard gate。

(c). Alice 對$2 \times n$個量子位元值$Q_A$根據$2 \times n$個控制訊號$A$各別操作後產生欲傳送的量子訊號 $Q_T = \{q_{T,2n-1}, q_{T,2n-2}, \dots, q_{T,1}, q_{T,0}\}$，並且通過量子通訊傳送量子訊號$Q_T$(即圖 10 中的❶)給 Bob。

Bob 依據下列步驟量測量子訊號和傳送控制訊號：

(d). Bob 隨機產生$2 \times n$個控制訊號，表示式為$B = \{b_{2n-1}, b_{2n-2}, \dots, b_1, b_0\}$；其中，第 $i$ 個控制訊號$b_i$決定第 $i$ 個量子位元$q_{T,i}$是否操作 Hadamard 閘，每個量子位元有$\frac{1}{2}$的機率會被操作 Hadamard 閘。

(e). Bob 對$2 \times n$個量子位元值$Q_T$根據$2 \times n$個控制訊號$B$各別操作後產生量測量子訊號$Q_B = \{q_{B,2\times n-1}, q_{B,2\times n-2}, \dots, q_{B,1}, q_{B,0}\}$，並且通過經典通訊傳送控制訊號$B$ (即圖 10 中的❷)給 Alice。

Alice 依據下列步驟比對雙方的控制訊號和回傳比對結果：

(f). Alice 接收到 Bob 的控制訊號$B$後比對 Alice 的控制訊號 $A$，產生比對結果 $C =$

$\{c_{2n-1}, c_{2n-2}, \dots, c_1, c_0\}$, where $c_i = \begin{cases} 1, a_i = b_i \\ 0, a_i \neq b_i \end{cases}$，並且通過經典通訊傳送比對結果$C$ (即圖 10 中的❸)給 Bob。除此之外，Alice 可以根據比對結果$C$取得具有相同控制訊號(即$c_i = 1$)的量子位元值作為協商後的結果。

Bob 依據下列步驟根據比對結果取得協商後的結果：

(g). Bob 可以根據比對結果$C$取得具有相同控制訊號(即$c_i = 1$)的量子位元值作為協商後的結果。

值得注意的是，雖然本節以到$2 \times n$個量子位元為例來說明，但考慮量子計算硬體成本也可以採用 1 個量子位元重覆執行 BB84 協定$2 \times n$次。

## 4.2 E91協定

在 E91 協定[13]基礎上，當 Alice 和 Bob 想協商$n$個位元的訊息(例如：$n$個位元的共享秘密金鑰)，則需要用到$3 \times n$個量子位元來傳輸。假設 Alice 為發起端，而 Bob 為接收端，E91 協定詳細流程如圖 11 所示，具體步驟描述如下。在此過程中，假設 Alice 與 Bob 共享三種控制訊號操作$\{s_1, s_2, s_3\}$，其中$s_1$代表不進行操作，$s_2$代表操作 Hadamard 閘，$s_3$代表先操作 S 閘再操作 Hadamard 閘。

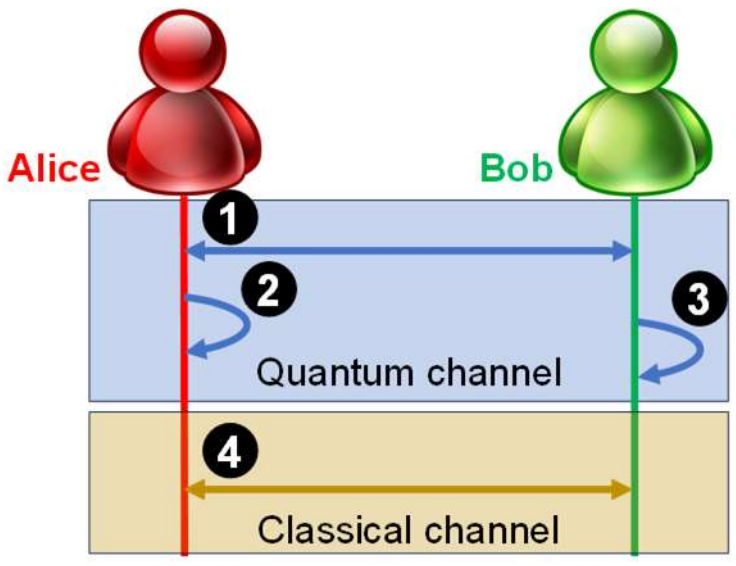

**圖 11 E91 協定**

在 E91 協定中，圖 11 中的訊息❶、❷、❸會循環執行$3 \times n$次。以下以第1次循環為例，說明每則訊息的細節如下：

(a). 在圖11的訊息❶中，會生成一對貝爾態的量子位元，之後Alice和Bob各持有其中一個量子位元。假設Alice持有量子位元$|q_j\rangle$，而 Bob 持有量子位元$|q_i\rangle$。

(b). 在圖11的訊息❷中，Alice隨機選擇三種控制訊號之一，並將所選控制訊號$a_l$施加於其量子位元$|q_j\rangle$。同樣地，在圖11的訊息❸中，Bob 隨機選擇三種控制訊號之一，並將所選控制訊號$b_l$施加於其量子位元$|q_i\rangle$。雙方施加控制訊號後，分別測量各自的量子態，Alice得到量子態 $q_{A,l}$，Bob得到量子態$q_{B,l}$。接著，流程回到訊息 的動作並重複，直到完成全部$3 \times n$次循環。

完成所有循環後，Alice 和 Bob 各自記錄$3 \times n$次迭代的控制訊號序列與測量結果。其中，Alice 獲得控制訊號序列$A' = \{a_{3n}, a_{3n-1}, \dots, a_2, a_1\}$，以及測量結果$Q_A' = \{q_{A,3n}, q_{A,3n-1}, \dots, q_{A,2}, q_{A,1}\}$。Bob 獲得控制訊號序列$B' = \{b_{3n}, b_{3n-1}, \dots, b_2, b_1\}$，以及測量結果$Q_B' = \{q_{B,3n}, q_{B,3n-1}, \dots, q_{B,2}, q_{B,1}\}$。

(c). 在圖11的訊息❹中，Alice和Bob通過經典通道交換各自的控制訊號序列，並進行比較以判斷是否匹配。比較結果為序列 $C' = \{c_{3n}, c_{3n-1}, \dots, c_2, c_1\}$, where $c_i = \begin{cases} 1, a_i = b_i \\ 0, a_i \neq b_i \end{cases}$。最後，提取對應於匹配控制訊號($c_i = 1$)的量子位元值，形成最終的共享秘密金鑰。

需要注意的是，當 Alice 和 Bob 都選擇$s_1$或$s_2$時，測量結果將會相同；但若雙方都選擇$s_3$，測量結果將會相反。在此情況下，Bob 必須對這些特定量子位元額外施加 X 閘操作以確保一致性。

## 4.3 訊息固定長度處理方法

本節將說明把協商後訊息做填補或取子 bit string 的方式來產生固定 $n$-bit 長度的訊息。為讓說明一下，本節假設BB84協定每次只傳一個Qubit，然後循環$2 \times n$次；E91 協定也是每次只傳一個 Qubit，然後循環$3 \times n$次。

由於第 4.1 節所述之 BB84 協定，通訊雙方主要有兩種控制訊號，所以循環$2 \times n$次後，平均可以得到 $n$ bits 可用資訊，詳細證明請見附錄 A。然而，每次協商後的訊息有可能少於 $n$ bits，也可能多於$n$ bits。同理，由於第 4.2 節所述之 E91 協定，通訊雙方主要有三種控制訊號，所以循環$3 \times n$次後，平均可以得到 $n$ bits 可用資訊，詳細證明請見附錄 B。然而，每次協商後的訊息仍有可能少於 $n$ bits，也可能多於 $n$ bits。

因此，本研究在訊息固定長度處理方法是作法如下：

- 當協商後的訊息少於 $n$ bits，則在 bit string 前面補 0，直到符合 $n$-bit 長度。
- 當協商後的訊息等於 $n$ bits，不做其他處理。
- 當協商後的訊息多於 $n$ bits，則只取 bit string 第 1 個位元到第 $n$ 個位元值。

# 5. 實驗結果與討論

本節將對本研究設計的量子金鑰分配與後量子密碼學混合方案的金鑰交換、量子金鑰分配與後量子密碼學混合方案的數位簽章進行效能驗證。首先，第 5.1 節說明實驗環境和相關設置。第 5.2 節和第 5.3 節驗證金鑰交換後得到的共享秘密金鑰的隨機位元，以及第 5.4 節分析各種金鑰交換協定的計算時間。最後，第 5.5 節討論量子金鑰分配與後量子密碼學混合方案的數位簽章應用在憑證的長度。

### 5.1 實驗環境

由於 GSM 協會(Groupe Speciale Mobile Association, GSMA)在 2024 年年底時發佈白皮書「IG.18 Opportunities and Challenges for Hybrid (QKD and PQC) Scenarios」，內容指出未來在核心網路設備有可能採用量子金鑰分配與後量子密碼學混合方案[6]。有鑑於此，本研究主要考量一般電腦的硬體環境，在硬體的部分主要採用 CPU Intel(R) Core(TM) i7-10510U、RAM 16.0 GB 進行相關實作。在軟體的部分，由於本研究主要聚焦在概念性驗證(Proof of Concept, POC)，並未實作量子晶片，所以量子計算主要建構在 IBM Qiskit SDK 1.1.1 以模擬器(simulator)方式實作。需要注意的是，未來仍需採用真實的量子晶片，才能具備量子物理特性和具備安全性。後量子密碼學演算法主要採用 BouncyCastle Open Source API 1.81 進行實作。

### 5.2 共享秘密金鑰熵驗證

本研究參考 NIST SP 800-90B 所述驗證方法[15]驗證共享秘密金鑰熵(entropy)，但受限於 ML-KEM 每次的共享秘密金鑰為 256-bit 長度，所以無法完全按照 NIST SP 800-90B 指定的 1000×1000 大小的 restart 矩陣。在本研究的作法主要讓每個金鑰交換協定各執行 1000 次，每次得到 256-bit 長度的共享秘密金鑰，所以 restart 矩陣的大小是 1000×256。之後計算 restart 矩陣的每一列的 0 和 1 發生的次數，並且取得最高次數作為 Most Common Value (MCV)，再根據 MCV 換算為最小熵(mini_entropy)和運用二項式分布檢定得到 $p$-value。

為公平比較，本研究主要比較 ECDH NIST P-256、ECDH Brainpool P-256、ML-KEM-512、QKD BB84、以及 QKD E91，共享秘密金鑰熵驗證結果如圖 12 所示。其中，由於每一列都是 256 個隨機位元值，所以如果 0 和 1 在均勻分布下期望值為各 128 次。因此，當 MCV 越接近 128，則最小熵(mini_entropy)值越大，以及 $p$-value 也會越大。當 $p$-value 低於門檻值時，則表示 0 和 1 不是均勻分布，混亂程度太低；該門檻值在 NIST SP 800-90B [15]中定義為 0.000005。取 MCV 觀察的用意在於表示該組資料是 1000 組資料中熵值最小的那組資料，也就是 worst case；當如果 worst case 都有驗證通過，剩下的 999 組資料也會通過。圖 12 採用盒鬚圖的方式呈現各種金鑰交換協定在各自 1000 組資料對應的 $p$-value，從分布可以觀察到全部都有高於門檻值 0.000005。表示各種金鑰交換協定可以產製足夠混亂的隨機位元，通過驗證。

由於基於 KMAC 的金鑰衍生函數是已被正式認可的金鑰衍生函數，所以如果輸入的資料源有通過熵驗證，則輸出的資料值也會通過熵驗證。因此，ECDH NIST P-256、ECDH Brainpool P-256、ML-KEM-512、QKD BB84、以及 QKD E91 得到的共享秘密金鑰有通過熵驗證，則這些金鑰交換協定的混合方案搭配基於 KMAC 的金鑰衍生函數產製的共享秘密金鑰也會通過熵驗證。

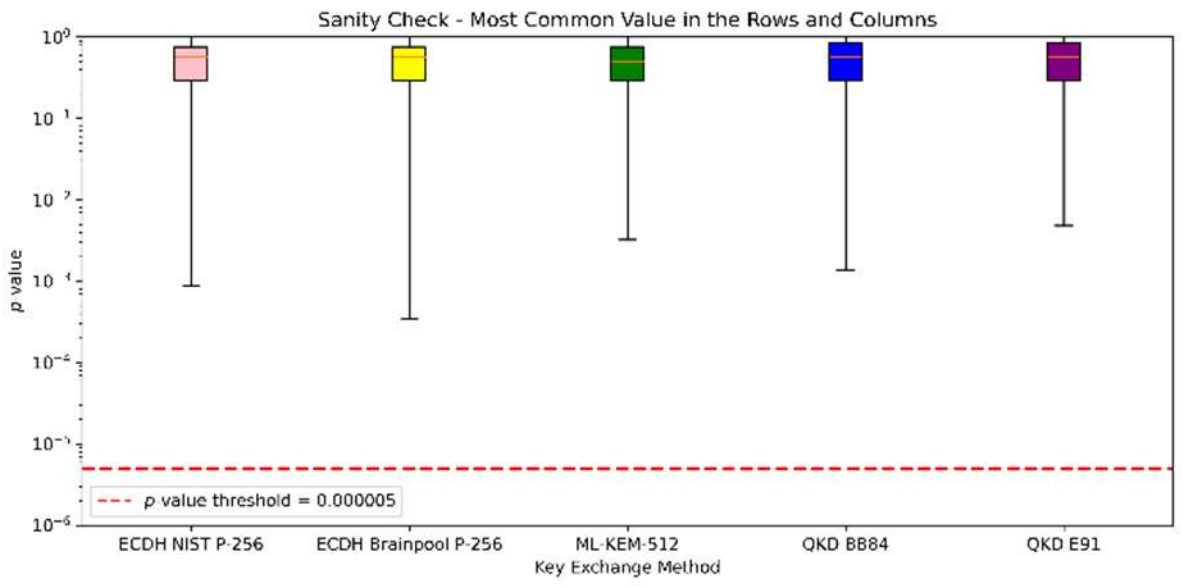

**圖 12 共享秘密金鑰熵驗證**

值得注意的是，QKD BB84 和 QKD E91 協商後得到的訊息長度有可能少於 $n$ bits，也可能多於 $n$ bits。雖然通過第 4.3 節的作法可以產生固定 $n$-bit 長度訊息，但在原本訊息少於 $n$ bits，採用填補 0 的方式，將造成 0 的位元數過多，導致熵驗證失敗。有鑑於此，本研究在 QKD BB84 和 QKD E91 的 $n$ 值設定為 384，讓協商後訊息值盡可能多於 $n$ bits，然後再前 $n$ bits 的值，以保障隨機性。

### 5.3 共享秘密金鑰獨立且同分布驗證

除了驗證量子隨機數產生器可以產製足夠混亂的隨機位元外，還需驗證是否為獨立且同分布(IID)。因此，NIST SP 800-90B 定義了獨立性檢定(Independence Test, Ind. Test)、適應度檢定(Goodness-of-fit Test, GF Test)、以及最長重覆子字串長度檢定(Length of the Longest Repeated Substring Test, Length of the LRS Test)[15]，但受限於 ML-KEM 每次的共享秘密金鑰為 256-bit 長度，所以本研究有基於原本的驗證精神行微調，分述如下。

(1). 獨立性檢定(Ind. Test)：第 5.2 節所述 restart 矩陣每一列都是 256 個隨機位元值，把每一組資料各別切割為 8 等份，每一等份裡各有 32 個隨機位元值，所以如果 0 和 1 在均勻分布下期望值為各 16 次。運用卡方檢定來驗證真值和期望值的平方差比例，並且加總後可以得每一組資料的卡方值，並且搭配自由度 7 可以計算得到 $p$-value，其中，$p$-value 門檻值在 NIST SP 800-90B [15]中定義為 0.001，對應的卡方值門檻值是 24.322。

(2). 適應度檢定(GF Test)：第 5.2 節所述 restart 矩陣每一列都是 256 個隨機位元值，觀察每 2 個位元的分布，把每一組資料各別切割為 128 等份，每一等份裡各有 2 個隨機位元值(即 00~11 的 4 種組合之一)，所以如果 00~11 在均勻分布下每一種組合的期望值為各 32 次。運用卡方檢定來驗證真值和期望值的平方差比例，並且加總後可以得每一組資料的卡方值，並且搭配自由度 2 可以計算得到 $p$-value，其中，$p$-value 門檻值在 NIST SP 800-90B [15]中定義為 0.001，對應的卡方值門檻值是 13.816。 (3). 最長重覆子字串長度檢定(Length of the LRS Test)：第 5.2 節所述 restart 矩陣每一列都是 256 個隨機位元值，觀察每一組資料中最長重覆子字串的長度，再運用二項式分布檢定計算 *p*-value。其中，*p*-value 門檻值在 NIST SP 800-90B [15]中定義為 0.001，所以在長度為 256 時對應的最長重覆子字串的長度門檻值大約為 24。

本研究主要比較 ECDH NIST P-256、ECDH Brainpool P-256、ML-KEM-512、QKD BB84、以及 QKD E91，共享秘密金鑰獨立且同分布驗證結果如圖 13、圖 14、圖 15 所示。其中，各個金鑰交換協定得到的共享秘密金鑰在獨立性檢定和適應度檢定的 *p*-value 以盒鬚圖呈現，分別如圖 13 和圖 14 所示。以及最長重覆子字串長度檢定的最長重覆子字串長度以盒鬚圖呈現，如圖 15 所示。由實驗結果可以觀察到即使在 worst case 的情況下，ML-KEM-512、QKD BB84、以及 QKD E91 的 *p*-value 也都有高於門檻值 0.001，所以 ML-KEM-512、QKD BB84、以及 QKD E91 都有通過獨立且同分布驗證。然而，ECDH NIST P-256 和 ECDH Brainpool P-256 卻有少數資料不符合獨立性驗證和適應性驗證。其原因可能是在於ECDH 協商後的共享秘密金鑰是取橢圓曲線點的 *x* 座標值，而橢圓曲線點的 *x* 座標值未完全符合獨立且同分布。

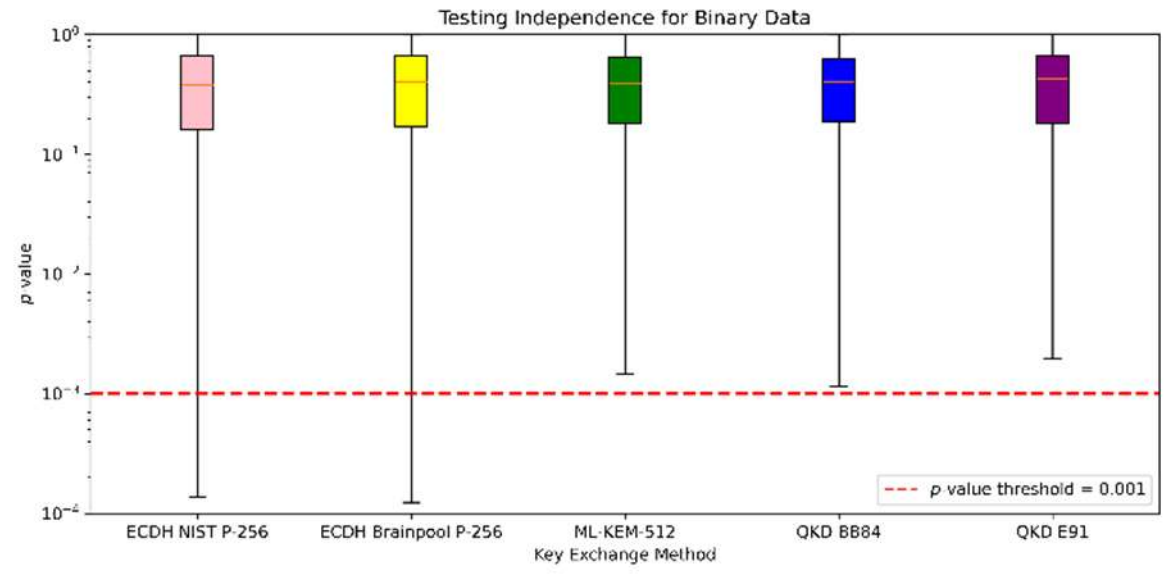

**圖 13 共享秘密金鑰獨立性驗證**

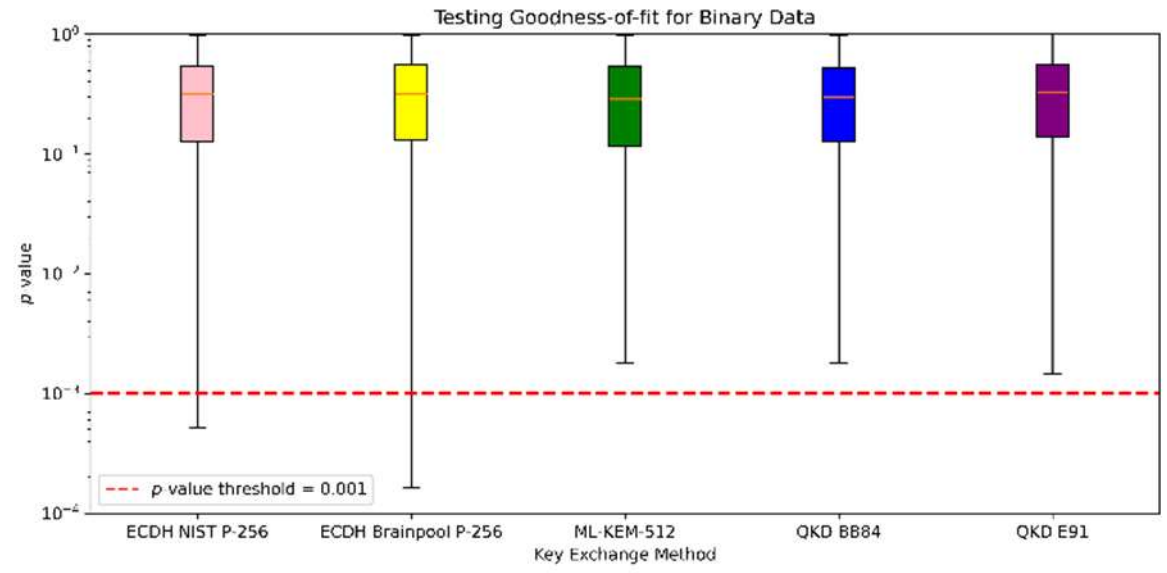

**圖 14 共享秘密金鑰適應性驗證**

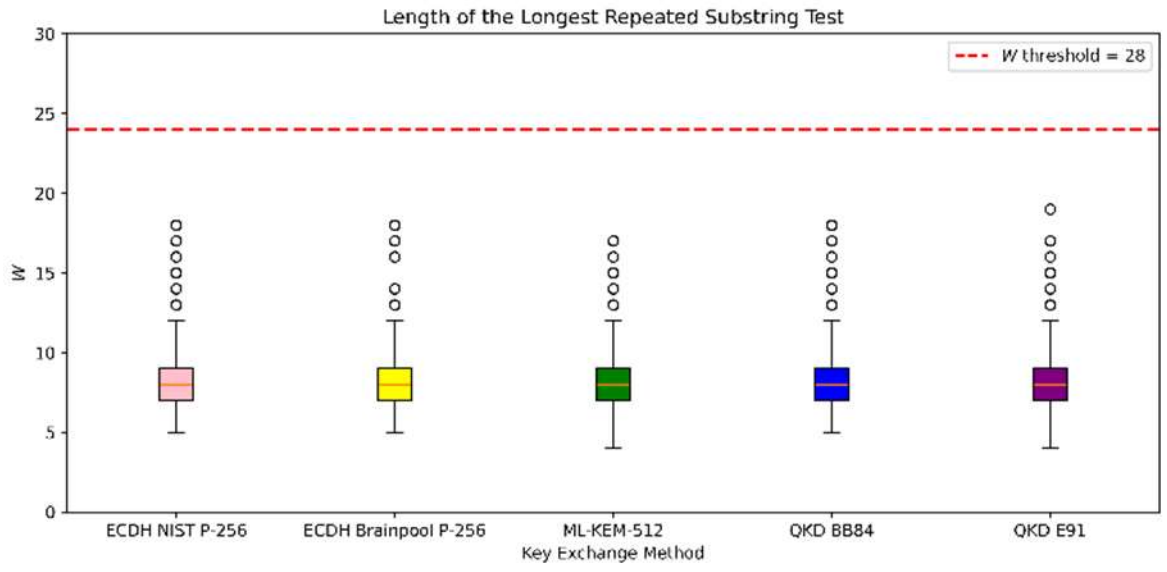

**圖 15 共享秘密金鑰最長子字串長度驗證**

## 5.4 計算時間比較

本研究主要比較 ECDH NIST、ECDH Brainpool、ML-KEM、QKD BB84、以及 QKD E91，各種金鑰交換協定的計算時間比較如圖 16 所示。

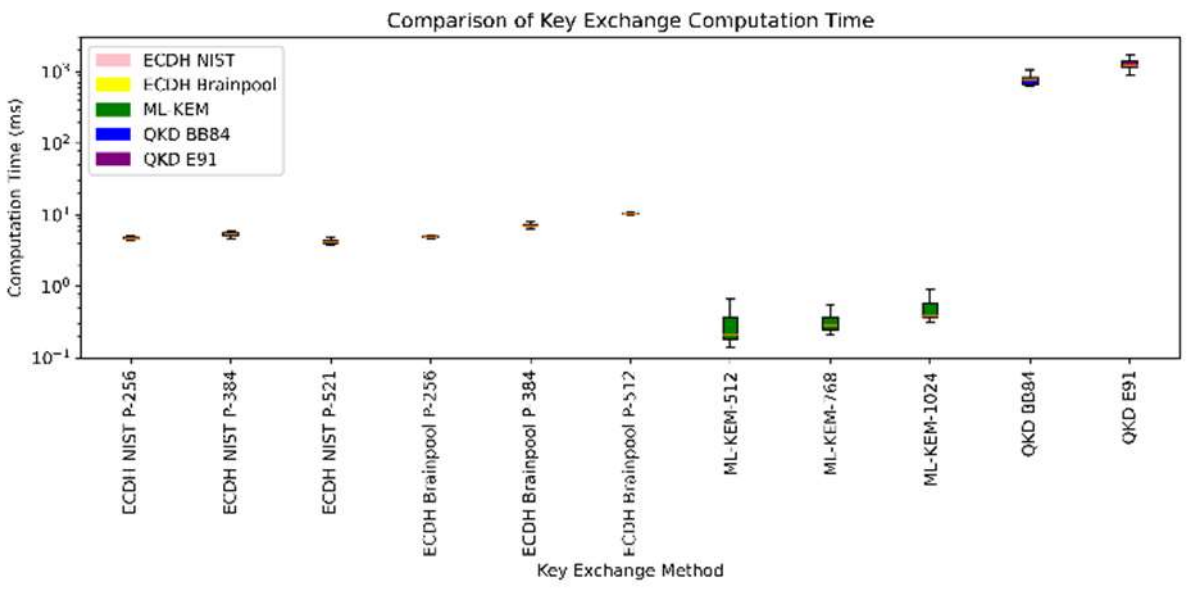

**圖 16 金鑰交換計算時間比較**

由第 2.1.1 節描述可知，基於 ECDH NIST 和 ECDH Brainpool 的金鑰交換協定主要需要計算 3 個橢圓曲線點的時間，包含產製 Alice 公鑰、產製 Bob 公鑰、以及協商後的橢圓曲線點，而橢圓曲線密碼學需要較多計算時間的也是在於橢圓曲線點的計算，所以 ECDH 需要較多的計算時間。

由第 2.1.2 節描述可知，基於 ML-KEM 的金鑰交換協定主要需要產生 ML-KEM 金鑰對、ML-KEM 金鑰封裝、以及 ML-KEM 解封裝三個計算時間的總合。由於 NIST 評選出來的後量子密碼學金鑰封裝機制是兼具安全和效率的演算法，所以擁有很高的效能，由實驗結果可以觀察到 ML-KEM 可以提供最高的效率。

在 QKD 的部分，由於本研究採用 IBM Qiskit 模擬器，所以執行效率較受限。除此之外，本研究是採用每次傳輸 1 個位元，然後循環多次的方式，以及 *n* 值設定為 384 (如第 5.2 節所述)。因此，BB84 協定需要循環 768 次，而 E91 需要循環 1152 次，所以需要較多的計算時間。未來可以考慮一次傳輸更多的位元數來減少循環次數，以及應該採用真實的量子晶片來達到真正的量子物理特性。

圖 16 中雖未提供混合方案的金鑰交換計算時間，但由於混合方案主要是不同金鑰交換的組合，所以可以根據各別所需的時間進行加總就能估計出各種混合方案的金鑰交換計算時間。

值得注意的是，由於本研究採用 IBM Qiskit 模擬器，所以各種金鑰交換協定都是在本機端完成，所以計算時間皆不包含網路傳輸時間。未來應該部署到不同的網路環境和設備，同時測試在網路傳輸時間影響下的金鑰交換計算時間。

## 5.5 長度比較

本節驗證本研究提出的量子金鑰分配與後量子密碼學混合方案的數位簽章，各種後量子密碼學數位簽章標準演算法的簽章長度和本研究提出的簽章請求碼(RCS)比較結果如表 1 所示。由於目

前後量子密碼學數位簽章標準演算法(如：ML-DSA 和 SLH-DSA)的簽章長度都較長，但本研究提出的量子金鑰分配與後量子密碼學混合方案的數位簽章可以在憑證中只放簽章請求碼，可以有效減少憑證中的簽章欄位長度。除此之外，本研究亦設計流程(如：圖 9 所示)，讓設備搭配量子金鑰分配等計算後得到簽章值，可以達到量子金鑰分配與後量子密碼學的雙重保障。另外，未來可以考慮把本研究提出的流程(如：圖 9 所示)結合到線上憑證狀態協定(Online Certificate Status Protocol, OCSP)，提供結合量子金鑰分配的線上憑證狀態協定，動態驗證憑證。

表 1 THE LENGTH COMPARISON OF SIGNATURE AND RCS BY THIS STUDY (UNIT: BYTES)

| DSA | Signature Length | RCS (this study) |
|---|---|---|
| ML-DSA-44 | 2420 | 32 |
| ML-DSA-65 | 3309 | 32 |
| ML-DSA-87 | 4627 | 32 |
| SLH-DSA-SHA2-128f<br>SLH-DSA-SHAKE-128f | 17088 | 32 |
| SLH-DSA-SHA2-192f<br>SLH-DSA-SHAKE-192f | 35664 | 32 |
| SLH-DSA-SHA2-256f<br>SLH-DSA-SHAKE-256f | 49856 | 32 |

## 6. 結論與未來研究

為提供後量子密碼學和量子物理特性雙重安全保障，本研究提出量子金鑰分配與 NIST 後量子密碼學標準演算法的混合方案應用於金鑰交換和數位簽章。

在金鑰交換的設計上，本研究提出基於量子金鑰分配與後量子密碼學混合方案的金鑰交換協定，運用 ML-KEM 得到隨機數，以及量子金鑰分配得到另一隨機數，再運用金鑰衍生函數產製共享秘密金鑰。在實驗中，本研究參考 NIST SP 800-90B 所述驗證方法[15]驗證共享秘密金鑰熵、共享秘密金鑰獨立且同分布，並且通過驗證，未來可以把此設計結合到傳輸層安全性協定。

在數位簽章的設計上，本研究提出量子金鑰分配與後量子密碼學混合方案的數位簽章，在憑證僅存放簽章請求碼，再由終端設備通過量子金鑰分配取得隨機數，以及憑證中心通過簽章值、簽章請求碼、隨機數產製簽章重構值(SRV)後發送給終端設備。終端設備可以通過簽章重構值、簽章請求碼、隨機數產製簽章值，並且進行驗章。在實驗中，本研究討論憑證中的簽章欄位長度，可以證明本研究方法有效縮短憑證長度，未來可以把此設計結合到線上憑證狀態協定。

本研究旨在設計量子金鑰分配與 NIST 後量子密碼學標準演算法的混合方案。然而，本研究目前僅為概念性驗證(POC)，在量子計實作上採用的是 IBM Qiskit SDK，並且是用模擬器(simulator)的方式，所以僅是模擬量子計算，並不具備真實量子物理特性。建議未來應該改為採用真實量子計算晶片，來提供量子物理特性，以及通過 NIST SP 800-90B 所述驗證方法。除此之外，本研究未考慮網路傳輸所需花費的時間，未來應該架設在不同的網路環境及其設備進行實證分析。期望未來 NIST 對量子金鑰分配能提供一些安全性指引文件或規範，感謝 NIST 卓越的貢獻。

## 致謝

本研究的量子隨機數產生器在實作上主要採用 IBM Qiskit SDK，以及本研究的後量子密碼學演算法在實作上主要採用 BouncyCastle Open Source API，感謝 IBM 和 BouncyCastle 提供免費 SDK 和 API，讓本研究得以順利進行，致上誠摯的感激。

## 參考文獻

## 附錄 A. BB84協定量子態證明

本節將說明 BB84 協定在傳輸過程中各種控制訊號操作下的量子態變化。其中，在本研究中的 BB84 協定通訊雙方只有兩種控制訊號：控制訊號(1)為不做任何操作、控制訊號(2)為操作 Hadamard gate (H-gate)。

以傳輸一個位元值為例，將先由 Alice 隨機設定初始值為$|0\rangle$或$|1\rangle$，並且$|0\rangle$或$|1\rangle$的機率各 0.5。然後 Alice 隨機選擇控制訊號，並且由於只有兩種控制訊號，每種控制訊號被選擇的機率各 0.5。之後Bob隨機選擇控制訊號，並且由於只有兩種控制訊號，每種控制訊號被選擇的機率各 0.5，操作後再進行量測。

假設$|0\rangle$的量子態向量表示為$\begin{bmatrix}1\\0\end{bmatrix}$，而$|1\rangle$的量子態向量表示為$\begin{bmatrix}0\\1\end{bmatrix}$。當選擇控制訊號(1)為不做任何操作時，則邏輯表示式為$I=\begin{bmatrix}1&0\\0&1\end{bmatrix}$，而當選擇控制訊號(2)為操作 H-gate 時，則邏輯表示式為$H=\frac{1}{\sqrt{2}}\begin{bmatrix}1&1\\1&-1\end{bmatrix}$。以下分別從 Alice 和 Bob 選擇不同控制訊號的情境、Alice 和 Bob 選擇相同控制訊號的情境進行討論。

### 1) Alice 和 Bob 選擇不同控制訊號的情境

假設 Alice 選擇控制訊號(1)，並且 Bob 選擇控制訊號(2)時，則隨著初始值為$|0\rangle$或$|1\rangle$可以得到結果如公式(A1)和公式(A2)所示。Bob 將得到均勻疊加態，不一定能量測到與 Alice 設定初始值一致的值。

$$HI|0\rangle=\frac{1}{\sqrt{2}}\begin{bmatrix}1&1\\1&-1\end{bmatrix}\begin{bmatrix}1&0\\0&1\end{bmatrix}\begin{bmatrix}1\\0\end{bmatrix}=\frac{1}{\sqrt{2}}\begin{bmatrix}1\\1\end{bmatrix}. \quad \text{(A1)}$$

$$HI|1\rangle=\frac{1}{\sqrt{2}}\begin{bmatrix}1&1\\1&-1\end{bmatrix}\begin{bmatrix}1&0\\0&1\end{bmatrix}\begin{bmatrix}0\\1\end{bmatrix}=\frac{1}{\sqrt{2}}\begin{bmatrix}1\\-1\end{bmatrix}. \quad \text{(A2)}$$

假設 Alice 選擇控制訊號(2)，並且 Bob 選擇控制訊號(1)時，則隨著初始值為$|0\rangle$或$|1\rangle$可以得到結果如公式(A3)和公式(A4)所示。Bob 也將得到均勻疊加態，不一定能量測到與 Alice 設定初始值一致的值。

$$IH|0\rangle=\begin{bmatrix}1&0\\0&1\end{bmatrix}\frac{1}{\sqrt{2}}\begin{bmatrix}1&1\\1&-1\end{bmatrix}\begin{bmatrix}1\\0\end{bmatrix}=\frac{1}{\sqrt{2}}\begin{bmatrix}1\\1\end{bmatrix}. \quad \text{(A3)}$$

$$IH|1\rangle=\begin{bmatrix}1&0\\0&1\end{bmatrix}\frac{1}{\sqrt{2}}\begin{bmatrix}1&1\\1&-1\end{bmatrix}\begin{bmatrix}0\\1\end{bmatrix}=\frac{1}{\sqrt{2}}\begin{bmatrix}1\\-1\end{bmatrix}. \quad \text{(A4)}$$

由上述結果可知，當 Alice 和 Bob 選擇不同控制訊號，則 Bob 量測到的位元值可能與 Alice 的位元值不一致，所以這些位元值將不被使用。

### 2) Alice 和 Bob 選擇相同控制訊號的情境

假設 Alice 選擇控制訊號(1)，並且 Bob 選擇控制訊號(1)時，則隨著初始值為$|0\rangle$或$|1\rangle$可以得到結果如公式(A5)和公式(A6)所示。Bob 將能量測到與 Alice 設定初始值一致的值。

$$II|0\rangle=|0\rangle. \quad \text{(A5)}$$

$$II|1\rangle=|1\rangle. \quad \text{(A6)}$$

假設 Alice 選擇控制訊號(2)，並且 Bob 選擇控制訊號(2)時，由於$HH=\begin{bmatrix}1&0\\0&1\end{bmatrix}$(如公式(A7)所示)，則隨著初始值為$|0\rangle$或$|1\rangle$可以得到結果如公式(A8)和公式(A9)所示。Bob 也將能量測到與 Alice 設定初始值一致的值。

$$\begin{aligned}HH&=\frac{1}{\sqrt{2}}\begin{bmatrix}1&1\\1&-1\end{bmatrix}\frac{1}{\sqrt{2}}\begin{bmatrix}1&1\\1&-1\end{bmatrix}\\&=\frac{1}{2}\begin{bmatrix}2&0\\0&2\end{bmatrix}=\begin{bmatrix}1&0\\0&1\end{bmatrix}.\end{aligned} \quad \text{(A7)}$$

$$HH|0\rangle=\begin{bmatrix}1&0\\0&1\end{bmatrix}\begin{bmatrix}1\\0\end{bmatrix}=\begin{bmatrix}1\\0\end{bmatrix}=|0\rangle. \quad \text{(A8)}$$

$$HH|1\rangle=\begin{bmatrix}1&0\\0&1\end{bmatrix}\begin{bmatrix}0\\1\end{bmatrix}=\begin{bmatrix}0\\1\end{bmatrix}=|1\rangle. \quad \text{(A9)}$$

由上述結果可知，當 Alice 和 Bob 選擇相同控制訊號，則 Bob 量測到的位元值可能與 Alice 的位元值一致，所以這些位元值可以被使用。

除此之外，由於 Alice 和 Bob 選擇控制訊號的排列組合展開共有 4 種可能，其中有 2 種可能可以量測到一致的值，所以當傳送$2n$位元時，平均有$n$位元值可以使用。

在安全性上，由於攻擊者在不知道 Alice 和 Bob 各自選擇的控制訊號的情況下，如果對量測態進行量測，將造成量子態塌陷，並且攻擊者無法覆現原本的量子態。

## 附錄 B. E91協定量子態證明

本節將說明 E91 協定在傳輸過程中各種控制訊號操作下的量子態變化。其中，在本研究中的 E91 協定通訊雙方只有三種控制訊號：控制訊號$s_1$為不做任何操作、控制訊號$s_2$為操作 H-gate、控制訊號$s_3$為操作 S-gate 和 H-gate。

假設$|q_i\rangle$和$|q_j\rangle$處於初始值皆為$|0\rangle$，為建立貝爾態其操作方式為先對$|q_j\rangle$操作 H-gate，再對$|q_iq_j\rangle$操作 CNOT-gate，並且控制位元是$|q_j\rangle$，邏輯表示

式為$C_x(|q_i\rangle \otimes H|q_j\rangle, j) = \frac{1}{\sqrt{2}}\begin{bmatrix}1\\0\\0\\1\end{bmatrix}$。由 Alice 持有處於貝爾態的$|q_j\rangle$，而 Bob 持有處於貝爾態的$|q_i\rangle$，再由 Alice 和 Bob 各自隨機選擇控制訊號，並且由於有三種控制訊號，每種控制訊號被選擇的機率各$\frac{1}{3}$，通訊雙方都操作後再進行量測。

當選擇控制訊號$s_1$為不做任何操作時，則邏輯表示式為$I = \begin{bmatrix}1 & 0\\0 & 1\end{bmatrix}$，當選擇控制訊號$s_2$為操作 H-gate 時，則邏輯表示式為$H = \frac{1}{\sqrt{2}}\begin{bmatrix}1 & 1\\1 & -1\end{bmatrix}$，當選擇控制訊號$s_3$為操作 S-gate 和 H-gate 時，則邏輯表示式為$HS = \frac{1}{\sqrt{2}}\begin{bmatrix}1 & 1\\1 & -1\end{bmatrix}\begin{bmatrix}1 & 0\\0 & \iota\end{bmatrix} = \frac{1}{\sqrt{2}}\begin{bmatrix}1 & \iota\\1 & -\iota\end{bmatrix}$。其中，$\iota$表示為虛數。以下分別從 Alice 和 Bob 選擇不同控制訊號的情境、Alice 和 Bob 選擇相同控制訊號的情境進行討論。

**1) Alice 選擇控制訊號 *s*1、Bob 選擇控制訊號 *s*1 的情境**

假設 Alice 選擇控制訊號$s_1$，並且 Bob 選擇控制訊號$s_1$時，則表示雙方選擇的控制訊號組合影響下的邏輯表示式為$I \otimes I$，如公式(B1)所示。因此，對原本處於貝爾態的$|q_i q_j\rangle$，在操作控制訊號組合後可以得到$\frac{1}{\sqrt{2}}(|00\rangle + |11\rangle)$，如公式(B2)所示。因此，Alice 和 Bob 一定可以量測到一致的值，並且量測到$|0\rangle$或$|1\rangle$的機率各為 0.5。因此，這個情境的量測結果可以採用。

$$
\begin{aligned}
I \otimes I &= \begin{bmatrix}1 & 0\\0 & 1\end{bmatrix} \otimes \begin{bmatrix}1 & 0\\0 & 1\end{bmatrix} \\
&= \begin{bmatrix} 1 \times \begin{bmatrix}1 & 0\\0 & 1\end{bmatrix} & 0 \times \begin{bmatrix}1 & 0\\0 & 1\end{bmatrix} \\ 0 \times \begin{bmatrix}1 & 0\\0 & 1\end{bmatrix} & 1 \times \begin{bmatrix}1 & 0\\0 & 1\end{bmatrix} \end{bmatrix} \\
&= \begin{bmatrix}1 & 0 & 0 & 0\\0 & 1 & 0 & 0\\0 & 0 & 1 & 0\\0 & 0 & 0 & 1\end{bmatrix}.
\end{aligned} \tag{B1}
$$

$$
\begin{aligned}
(I \otimes I)C_x(|q_i\rangle \otimes H|q_j\rangle, j) &= \begin{bmatrix}1 & 0 & 0 & 0\\0 & 1 & 0 & 0\\0 & 0 & 1 & 0\\0 & 0 & 0 & 1\end{bmatrix} \frac{1}{\sqrt{2}}\begin{bmatrix}1\\0\\0\\1\end{bmatrix} \\
&= \frac{1}{\sqrt{2}}\begin{bmatrix}1\\0\\0\\1\end{bmatrix} = \frac{1}{\sqrt{2}}(|00\rangle + |11\rangle).
\end{aligned} \tag{B2}
$$

**2) Alice 選擇控制訊號 *s*1、Bob 選擇控制訊號 *s*2 的情境**

假設 Alice 選擇控制訊號$s_1$，並且 Bob 選擇控制訊號$s_2$時，則表示雙方選擇的控制訊號組合影響下的邏輯表示式為$H \otimes I$，如公式(B3)所示。因此，對原本處於貝爾態的$|q_i q_j\rangle$，在操作控制訊號組合後可以得到$\frac{1}{2}(|00\rangle + |01\rangle + |10\rangle - |11\rangle)$，如公式(B4)所示。因此，Alice 和 Bob 不一定可以量測到一致的值，量測到的值不一致的機率為 0.5。因此，這個情境的量測結果無法採用。

$$
\begin{aligned}
H \otimes I &= \frac{1}{\sqrt{2}}\begin{bmatrix}1 & 1\\1 & -1\end{bmatrix} \otimes \begin{bmatrix}1 & 0\\0 & 1\end{bmatrix} \\
&= \frac{1}{\sqrt{2}}\begin{bmatrix} 1 \times \begin{bmatrix}1 & 0\\0 & 1\end{bmatrix} & 1 \times \begin{bmatrix}1 & 0\\0 & 1\end{bmatrix} \\ 1 \times \begin{bmatrix}1 & 0\\0 & 1\end{bmatrix} & (-1) \times \begin{bmatrix}1 & 0\\0 & 1\end{bmatrix} \end{bmatrix} \\
&= \frac{1}{\sqrt{2}}\begin{bmatrix}1 & 0 & 1 & 0\\0 & 1 & 0 & 1\\1 & 0 & -1 & 0\\0 & 1 & 0 & -1\end{bmatrix}.
\end{aligned} \tag{B3}
$$

$$
\begin{aligned}
(H \otimes I)C_x(|q_i\rangle \otimes H|q_j\rangle, j) &= \frac{1}{\sqrt{2}}\begin{bmatrix}1 & 0 & 1 & 0\\0 & 1 & 0 & 1\\1 & 0 & -1 & 0\\0 & 1 & 0 & -1\end{bmatrix} \frac{1}{\sqrt{2}}\begin{bmatrix}1\\0\\0\\1\end{bmatrix} \\
&= \frac{1}{2}\begin{bmatrix}1\\1\\1\\-1\end{bmatrix} = \frac{1}{2}(|00\rangle + |01\rangle + |10\rangle \\
&\qquad - |11\rangle).
\end{aligned} \tag{B4}
$$

**3) Alice 選擇控制訊號 *s*1、Bob 選擇控制訊號 *s*3 的情境**

假設 Alice 選擇控制訊號$s_1$，並且 Bob 選擇控制訊號$s_3$時，則表示雙方選擇的控制訊號組合影響下的邏輯表示式為$HS \otimes I$，如公式(B5)所示。因此，對原本處於貝爾態的$|q_i q_j\rangle$，在操作控制訊號組合後可以得到$\frac{1}{2}(|00\rangle + \iota|01\rangle + |10\rangle - \iota|11\rangle)$，如公式(B6)所示。因此，Alice 和 Bob 不一定可以量測到一致的值，量測到的值不一致的機率為 0.5。因此，這個情境的量測結果無法採用。

$$
\begin{aligned}
(HS) \otimes I &= \frac{1}{\sqrt{2}}\begin{bmatrix}1 & \iota\\1 & -\iota\end{bmatrix} \otimes \begin{bmatrix}1 & 0\\0 & 1\end{bmatrix} \\
&= \frac{1}{\sqrt{2}}\begin{bmatrix} 1 \times \begin{bmatrix}1 & 0\\0 & 1\end{bmatrix} & \iota \times \begin{bmatrix}1 & 0\\0 & 1\end{bmatrix} \\ 1 \times \begin{bmatrix}1 & 0\\0 & 1\end{bmatrix} & (-\iota) \times \begin{bmatrix}1 & 0\\0 & 1\end{bmatrix} \end{bmatrix} \\
&= \frac{1}{\sqrt{2}}\begin{bmatrix}1 & 0 & \iota & 0\\0 & 1 & 0 & \iota\\1 & 0 & -\iota & 0\\0 & 1 & 0 & -\iota\end{bmatrix}.
\end{aligned} \tag{B5}
$$

$$
\begin{aligned}
((HS) \otimes I)C_x(|q_i\rangle \otimes H|q_j\rangle, j) &= \frac{1}{\sqrt{2}}\begin{bmatrix}1 & 0 & \iota & 0\\0 & 1 & 0 & \iota\\1 & 0 & -\iota & 0\\0 & 1 & 0 & -\iota\end{bmatrix} \frac{1}{\sqrt{2}}\begin{bmatrix}1\\0\\0\\1\end{bmatrix} \\
&= \frac{1}{2}\begin{bmatrix}1\\\iota\\1\\-\iota\end{bmatrix} = \frac{1}{2}(|00\rangle + \iota|01\rangle + |10\rangle \\
&\qquad - \iota|11\rangle).
\end{aligned} \tag{B6}
$$

**4) Alice 選擇控制訊號 $s_2$、Bob 選擇控制訊號 $s_1$ 的情境**

假設 Alice 選擇控制訊號$s_2$，並且 Bob 選擇控制訊號$s_1$時，則表示雙方選擇的控制訊號組合影響下的邏輯表示式為$I \otimes H$，如公式(B7)所示。因此，對原本處於貝爾態的$|q_i q_j\rangle$，在操作控制訊號組合後可以得到$\frac{1}{2}(|00\rangle+|01\rangle+|10\rangle-|11\rangle)$，如公式(B8)所示。因此，Alice 和 Bob 不一定可以量測到一致的值，量測到的值不一致的機率為 0.5。因此，這個情境的量測結果無法採用。

$$\begin{aligned} I\otimes H &= \begin{bmatrix}1&0\\0&1\end{bmatrix}\otimes\frac{1}{\sqrt{2}}\begin{bmatrix}1&1\\1&-1\end{bmatrix} \\ &= \frac{1}{\sqrt{2}}\begin{bmatrix}1\times\begin{bmatrix}1&1\\1&-1\end{bmatrix} & 0\times\begin{bmatrix}1&1\\1&-1\end{bmatrix}\\ 0\times\begin{bmatrix}1&1\\1&-1\end{bmatrix} & 1\times\begin{bmatrix}1&1\\1&-1\end{bmatrix}\end{bmatrix} \\ &= \frac{1}{\sqrt{2}}\begin{bmatrix}1&1&0&0\\1&-1&0&0\\0&0&1&1\\0&0&1&-1\end{bmatrix}. \end{aligned} \tag{B7}$$

$$\begin{aligned} (I\otimes H)C_x(|q_i\rangle\otimes H|q_j\rangle, j) &= \frac{1}{\sqrt{2}}\begin{bmatrix}1&1&0&0\\1&-1&0&0\\0&0&1&1\\0&0&1&-1\end{bmatrix}\frac{1}{\sqrt{2}}\begin{bmatrix}1\\0\\0\\1\end{bmatrix} \\ &= \frac{1}{2}\begin{bmatrix}1\\1\\1\\-1\end{bmatrix} = \frac{1}{2}(|00\rangle+|01\rangle+|10\rangle \\ &\quad -|11\rangle). \end{aligned} \tag{B8}$$

**5) Alice 選擇控制訊號 $s_2$、Bob 選擇控制訊號 $s_2$ 的情境**

假設 Alice 選擇控制訊號$s_2$，並且 Bob 選擇控制訊號$s_2$時，則表示雙方選擇的控制訊號組合影響下的邏輯表示式為$H \otimes H$，如公式(B9)所示。因此，對原本處於貝爾態的$|q_i q_j\rangle$，在操作控制訊號組合後可以得到$\frac{1}{\sqrt{2}}(|00\rangle+|11\rangle)$，如公式(B10)所示。因此，Alice 和 Bob 一定可以量測到一致的值，並且量測到$|0\rangle$或$|1\rangle$的機率各為 0.5。因此，這個情境的量測結果可以採用。

$$\begin{aligned} H\otimes H &= \frac{1}{\sqrt{2}}\begin{bmatrix}1&1\\1&-1\end{bmatrix}\otimes\frac{1}{\sqrt{2}}\begin{bmatrix}1&1\\1&-1\end{bmatrix} \\ &= \frac{1}{2}\begin{bmatrix}1\times\begin{bmatrix}1&1\\1&-1\end{bmatrix} & 1\times\begin{bmatrix}1&1\\1&-1\end{bmatrix}\\ 1\times\begin{bmatrix}1&1\\1&-1\end{bmatrix} & (-1)\times\begin{bmatrix}1&1\\1&-1\end{bmatrix}\end{bmatrix} \\ &= \frac{1}{2}\begin{bmatrix}1&1&1&1\\1&-1&1&-1\\1&1&-1&-1\\1&-1&-1&1\end{bmatrix}. \end{aligned} \tag{B9}$$

$$\begin{aligned} (H\otimes H)C_x(|q_i\rangle\otimes H|q_j\rangle, j) &= \frac{1}{2}\begin{bmatrix}1&1&1&1\\1&-1&1&-1\\1&1&-1&-1\\1&-1&-1&1\end{bmatrix}\frac{1}{\sqrt{2}}\begin{bmatrix}1\\0\\0\\1\end{bmatrix} \\ &= \frac{1}{2\sqrt{2}}\begin{bmatrix}2\\0\\0\\2\end{bmatrix} = \frac{1}{\sqrt{2}}\begin{bmatrix}1\\0\\0\\1\end{bmatrix} \\ &= \frac{1}{\sqrt{2}}(|00\rangle+|11\rangle). \end{aligned} \tag{B10}$$

**6) Alice 選擇控制訊號 $s_2$、Bob 選擇控制訊號 $s_3$ 的情境**

假設 Alice 選擇控制訊號$s_2$，並且 Bob 選擇控制訊號$s_3$時，則表示雙方選擇的控制訊號組合影響下的邏輯表示式為$HS \otimes H$，如公式(B11)所示。因此，對原本處於貝爾態的$|q_i q_j\rangle$，在操作控制訊號組合後可以得到$\frac{1+\iota}{2\sqrt{2}}|00\rangle+\frac{1-\iota}{2\sqrt{2}}|01\rangle+\frac{1-\iota}{2\sqrt{2}}|10\rangle+\frac{1+\iota}{2\sqrt{2}}|11\rangle$，如公式(B12)所示。因此，Alice 和 Bob 不一定可以量測到一致的值，量測到的值不一致的機率為 0.5。因此，這個情境的量測結果無法採用。

$$\begin{aligned} (HS)\otimes H &= \frac{1}{\sqrt{2}}\begin{bmatrix}1&\iota\\1&-\iota\end{bmatrix}\otimes\frac{1}{\sqrt{2}}\begin{bmatrix}1&1\\1&-1\end{bmatrix} \\ &= \frac{1}{2}\begin{bmatrix}1\times\begin{bmatrix}1&1\\1&-1\end{bmatrix} & \iota\times\begin{bmatrix}1&1\\1&-1\end{bmatrix}\\ 1\times\begin{bmatrix}1&1\\1&-1\end{bmatrix} & (-\iota)\times\begin{bmatrix}1&1\\1&-1\end{bmatrix}\end{bmatrix} \\ &= \frac{1}{2}\begin{bmatrix}1&1&\iota&\iota\\1&-1&\iota&-\iota\\1&1&-\iota&-\iota\\1&-1&-\iota&\iota\end{bmatrix}. \end{aligned} \tag{B11}$$

$$\begin{aligned} ((HS)\otimes H)C_x(|q_i\rangle\otimes H|q_j\rangle, j) &= \frac{1}{2}\begin{bmatrix}1&1&\iota&\iota\\1&-1&\iota&-\iota\\1&1&-\iota&-\iota\\1&-1&-\iota&\iota\end{bmatrix}\frac{1}{\sqrt{2}}\begin{bmatrix}1\\0\\0\\1\end{bmatrix} \\ &= \frac{1}{2\sqrt{2}}\begin{bmatrix}1+\iota\\1-\iota\\1-\iota\\1+\iota\end{bmatrix} \\ &= \frac{1+\iota}{2\sqrt{2}}|00\rangle+\frac{1-\iota}{2\sqrt{2}}|01\rangle+\frac{1-\iota}{2\sqrt{2}}|10\rangle+ \\ &\quad \frac{1+\iota}{2\sqrt{2}}|11\rangle. \end{aligned} \tag{B12}$$

**7) Alice 選擇控制訊號 $s_3$、Bob 選擇控制訊號 $s_1$ 的情境**

假設 Alice 選擇控制訊號$s_3$，並且 Bob 選擇控制訊號$s_1$時，則表示雙方選擇的控制訊號組合影響下的邏輯表示式為$I \otimes HS$，如公式(B13)所示。因此，對原本處於貝爾態的$|q_i q_j\rangle$，在操作控制訊號組合後可以得到$\frac{1}{2}(|00\rangle+|01\rangle+\iota|10\rangle-\iota|11\rangle)$，如公式(B14)所示。因此，Alice 和 Bob 不一定可以量測到

一致的值，量測到的值不一致的機率為 0.5。因此，這個情境的量測結果無法採用。

$$
\begin{aligned}
&I\otimes(HS)\\
&=\begin{bmatrix}1&0\\0&1\end{bmatrix}\otimes\frac{1}{\sqrt{2}}\begin{bmatrix}1&\iota\\1&-\iota\end{bmatrix}\\
&=\frac{1}{\sqrt{2}}\begin{bmatrix}1\times\begin{bmatrix}1&\iota\\1&-\iota\end{bmatrix}&0\times\begin{bmatrix}1&\iota\\1&-\iota\end{bmatrix}\\0\times\begin{bmatrix}1&\iota\\1&-\iota\end{bmatrix}&1\times\begin{bmatrix}1&\iota\\1&-\iota\end{bmatrix}\end{bmatrix}\\
&=\frac{1}{\sqrt{2}}\begin{bmatrix}1&\iota&0&0\\1&-\iota&0&0\\0&0&1&\iota\\0&0&1&-\iota\end{bmatrix}.
\end{aligned}
\tag{B13}
$$

$$
\begin{aligned}
&\big(I\otimes(HS)\big)C_x\big(|q_i\rangle\otimes H|q_j\rangle,j\big)\\
&=\frac{1}{\sqrt{2}}\begin{bmatrix}1&\iota&0&0\\1&-\iota&0&0\\0&0&1&\iota\\0&0&1&-\iota\end{bmatrix}\frac{1}{\sqrt{2}}\begin{bmatrix}1\\0\\0\\1\end{bmatrix}\\
&=\frac{1}{2}\begin{bmatrix}1\\1\\\iota\\-\iota\end{bmatrix}=\frac{1}{2}(|00\rangle+|01\rangle+\iota|10\rangle\\
&\qquad-\iota|11\rangle).
\end{aligned}
\tag{B14}
$$

**8) Alice 選擇控制訊號 $s_3$、Bob 選擇控制訊號 $s_2$ 的情境**

假設 Alice 選擇控制訊號$s_3$，並且 Bob 選擇控制訊號$s_2$時，則表示雙方選擇的控制訊號組合影響下的邏輯表示式為$H\otimes HS$，如公式(B15)所示。因此，對原本處於貝爾態的$|q_iq_j\rangle$，在操作控制訊號組合後可以得到$\frac{1+\iota}{2\sqrt{2}}|00\rangle+\frac{1-\iota}{2\sqrt{2}}|01\rangle+\frac{1-\iota}{2\sqrt{2}}|10\rangle+\frac{1+\iota}{2\sqrt{2}}|11\rangle$，如公式(B16)所示。因此，Alice 和 Bob 不一定可以量測到一致的值，量測到的值不一致的機率為 0.5。因此，這個情境的量測結果無法採用。

$$
\begin{aligned}
&H\otimes(HS)\\
&=\frac{1}{\sqrt{2}}\begin{bmatrix}1&1\\1&-1\end{bmatrix}\otimes\frac{1}{\sqrt{2}}\begin{bmatrix}1&\iota\\1&-\iota\end{bmatrix}\\
&=\frac{1}{2}\begin{bmatrix}1\times\begin{bmatrix}1&\iota\\1&-\iota\end{bmatrix}&1\times\begin{bmatrix}1&\iota\\1&-\iota\end{bmatrix}\\1\times\begin{bmatrix}1&\iota\\1&-\iota\end{bmatrix}&(-1)\times\begin{bmatrix}1&\iota\\1&-\iota\end{bmatrix}\end{bmatrix}\\
&=\frac{1}{2}\begin{bmatrix}1&\iota&1&\iota\\1&-\iota&1&-\iota\\1&\iota&-1&-\iota\\1&-\iota&-1&\iota\end{bmatrix}.
\end{aligned}
\tag{B15}
$$

$$
\begin{aligned}
&\big((HS)\otimes(HS)\big)C_x\big(|q_i\rangle\otimes H|q_j\rangle,j\big)\\
&=\frac{1}{2}\begin{bmatrix}1&\iota&1&\iota\\1&-\iota&1&-\iota\\1&\iota&-1&-\iota\\1&-\iota&-1&\iota\end{bmatrix}\frac{1}{\sqrt{2}}\begin{bmatrix}1\\0\\0\\1\end{bmatrix}\\
&=\frac{1}{2\sqrt{2}}\begin{bmatrix}1+\iota\\1-\iota\\1-\iota\\1+\iota\end{bmatrix}=\frac{1}{\sqrt{2}}\begin{bmatrix}0\\1\\1\\0\end{bmatrix}\\
&=\frac{1+\iota}{2\sqrt{2}}|00\rangle+\frac{1-\iota}{2\sqrt{2}}|01\rangle+\frac{1-\iota}{2\sqrt{2}}|10\rangle+\\
&\qquad\frac{1+\iota}{2\sqrt{2}}|11\rangle.
\end{aligned}
\tag{B16}
$$

**9) Alice 選擇控制訊號 $s_3$、Bob 選擇控制訊號 $s_3$ 的情境**

假設 Alice 選擇控制訊號$s_3$，並且 Bob 選擇控制訊號$s_3$時，則表示雙方選擇的控制訊號組合影響下的邏輯表示式為$HS\otimes HS$，如公式(B17)所示。因此，對原本處於貝爾態的$|q_iq_j\rangle$，在操作控制訊號組合後可以得到$\frac{1}{\sqrt{2}}(|01\rangle+|10\rangle)$，如公式(B18)所示。因此，Alice 和 Bob 一定可以量測到相反的值，並且量測到$|0\rangle$或$|1\rangle$的機率各為 0.5。因此，這個情境，可以限定讓 Bob 量測到的變號($|0\rangle$變$|1\rangle$、$|1\rangle$變$|0\rangle$)即可，量測結果可以採用。

$$
\begin{aligned}
&(HS)\otimes(HS)\\
&=\frac{1}{\sqrt{2}}\begin{bmatrix}1&\iota\\1&-\iota\end{bmatrix}\otimes\frac{1}{\sqrt{2}}\begin{bmatrix}1&\iota\\1&-\iota\end{bmatrix}\\
&=\frac{1}{2}\begin{bmatrix}1\times\begin{bmatrix}1&\iota\\1&-\iota\end{bmatrix}&\iota\times\begin{bmatrix}1&\iota\\1&-\iota\end{bmatrix}\\1\times\begin{bmatrix}1&\iota\\1&-\iota\end{bmatrix}&(-\iota)\times\begin{bmatrix}1&\iota\\1&-\iota\end{bmatrix}\end{bmatrix}\\
&=\frac{1}{2}\begin{bmatrix}1&\iota&\iota&-1\\1&-\iota&\iota&1\\1&\iota&-\iota&1\\1&-\iota&-\iota&-1\end{bmatrix}.
\end{aligned}
\tag{B17}
$$

$$
\begin{aligned}
&\big((HS)\otimes(HS)\big)C_x\big(|q_i\rangle\otimes H|q_j\rangle,j\big)\\
&=\frac{1}{2}\begin{bmatrix}1&\iota&\iota&-1\\1&-\iota&\iota&1\\1&\iota&-\iota&1\\1&-\iota&-\iota&-1\end{bmatrix}\frac{1}{\sqrt{2}}\begin{bmatrix}1\\0\\0\\1\end{bmatrix}\\
&=\frac{1}{2\sqrt{2}}\begin{bmatrix}0\\2\\2\\0\end{bmatrix}=\frac{1}{\sqrt{2}}\begin{bmatrix}0\\1\\1\\0\end{bmatrix}\\
&=\frac{1}{\sqrt{2}}(|01\rangle+|10\rangle).
\end{aligned}
\tag{B18}
$$

**10) 討論**

由上述結果可知，當 Alice 和 Bob 選擇相同控制訊號，則 Bob 能量測到的位元值能與 Alice 的位元值一致，所以這些位元值可以被使用。

除此之外，由於 Alice 和 Bob 選擇控制訊號的排列組合展開共有 9 種可能，其中有 3 種可能可以量測到一致的值，所以當傳送$3n$位元時，平均有 $n$ 位元值可以使用。

在安全性上，由於攻擊者在不知道 Alice 和 Bob 各自選擇的控制訊號的情況下，由於$|q_iq_j\rangle$有可能存在$|00\rangle$、$|01\rangle$、$|10\rangle$、$|11\rangle$各 0.25 的機率，所以攻擊者無法肯定量測的值是否會與 Alice 量測的值一致。另外，如果對量測態進行量測，將造成量子態塌陷，並且攻擊者無法覆現原本的量子態。